\documentclass[letterpaper,twocolumn,10pt]{article}
\usepackage{usenix2019,epsfig,endnotes}

\usepackage[english]{babel}
\usepackage[utf8x]{inputenc}
\usepackage[T1]{fontenc}

\usepackage{amsmath}
\usepackage{graphicx}
\usepackage[colorinlistoftodos]{todonotes}

\usepackage{algorithm}
\usepackage[noend]{algpseudocode}

\usepackage[normalem]{ulem}
\usepackage{units}
\usepackage{verbatim}

\usepackage[super]{nth}
\usepackage{lscape}
\usepackage{multirow}
\usepackage{hyperref}

\usepackage{graphicx}
\graphicspath{ {.} }

\newif\iflong
\longtrue

\DeclareMathOperator{\rank}{rank}
\DeclareMathOperator{\Num}{Num}
\DeclareMathOperator{\Vectorize}{Vec}

\newcommand{\remove}[1]{}
\newcommand{\IPID}{\mathit{IPID}}
\newcommand{\IP}{\mathit{IP}}
\newcommand{\itemshort}{\vspace{-0.25cm} \item}
\newcommand{\mypar}[1]{\vspace{-0.5cm}\paragraph{#1}}

\makeatletter
\renewcommand{\@maketitle}{%
    \newpage\null\vskip2em%
    \begin{center}%
        \let\footnote\thanks{\LARGE\@title\par}%
        \vskip1.5em{\large\lineskip.5em\begin{tabular}[t]{c}\@author\end{tabular}\par }%
        \vskip 1em{\large \@date }%
    \end{center}%
    \par}
\makeatother

\begin{document}

\date{}

\iflong
\title{\Large \bf   From IP ID to Device ID and KASLR Bypass (Extended Version)\thanks{This is an extended version of a paper that will be published in Usenix Security 2019.}
}
\fi
\author{
{\rm Amit Klein}\\
Bar-Ilan University
\and
{\rm Benny Pinkas}\\
Bar-Ilan University
} 

\maketitle

\pagestyle{empty}

\subsection*{Abstract} 
IP headers include a 16-bit ID field. Our work examines the  generation of this field in Windows (versions 8 and higher), Linux and Android,  and shows that the IP ID field enables remote servers to assign a unique ID to each  device and thus be able to identify subsequent transmissions sent from that device. This identification works across all browsers and over network changes. In modern Linux and Android versions, this field leaks a kernel address, thus we also break KASLR. 

Our work includes reverse-engineering of the Windows IP ID generation code, and a  cryptanalysis of this code and of the Linux kernel IP ID generation code. It provides practical techniques to partially extract the key used by each of these algorithms, overcoming different implementation issues, 
and observing that this key can identify individual devices. We deployed a demo (for Windows) showing that key extraction and machine fingerprinting works in the wild, and
tested it from networks around the world. 
\section{Introduction}
\label{sec:intro}
Online browser-based user tracking is prevalent. Tracking is used to identify users and track them across many sessions and websites on the Internet. Tracking is often performed in order to  personalize advertisements or for surveillance purposes. It can either be done by sites that are visited by users, or by third-party companies which track users across multiple web sites and applications. \cite{Acar:2013:FDW:2541806.2516674} specifically lists motivations for web-based fingerprinting as ``fraud detection, protection against account hijacking, anti-bot and anti-scraping services, enterprise security management, protection against DDOS attacks, real-time targeted marketing, campaign measurement, reaching customers across devices, and limiting number of access to
services''.

\mypar{Tracking methods}
Existing tracking mechanisms are usually based on either {\em tagging} or {\em fingerprinting}. With tagging, the tracking party stores at the user's device some information, such as a cookie, which can later be tracked. Modern web standards and norms, however, enable users to opt-out from tagging. Furthermore, tagging is often specific for one application or browser, and therefore a tag that was stored in one browser cannot be identified when the user is using a different browser on the same machine, or when the user uses the private browsing feature of the browser.   Fingerprinting is implemented by having the tracking party measure features of the user's machine (for example the set of installed fonts). Corporates, however, often install a single ``golden image'' (standard set of software packages) on many {\em identical} (hardware-wise)  machines, and therefore it is hard to obtain fingerprints that distinguish among such machines. 

In this work we present a new tracking mechanism which is based on extracting data used by the IP ID generator (see Section~\ref{sec:ipid}). It is the first tracking technique that is able to simultaneously (a) cross the private browsing boundary
(i.e. compute  the same tracking ID for a private mode tab/window of a browser as for a regular tab/window of the browser); (b) work across different browsers; (c) address the ``golden image'' problem;  and (d) work across multiple  networks; all this while maintaining a very good coverage of the platforms involved.
To our knowledge, no other tracking method (or a combination of several tracking techniques) achieves all these goals simultaneously.
Moreover, the Windows variant of this technique also survives Windows shutdown+startup (but not restart).

Our techniques are realistic: for Windows we only need to have control over 8-30 IP addresses (in 3-13 class B networks), and for Linux/Android, we only need to control 300-400 IP addresses (can be in a single class B network). The Windows technique was successfully tested in the wild.

\subsection{Introduction to IP ID}
\label{sec:ipid}
The IP ID field is a 16 bit IP header field, defined in RFC 791 \cite{rfc791}. It is used to facilitate de-fragmentation, by marking IP fragments that belong to the same IP datagram. The IP protocol assembles fragments into a datagram based on the fragment source IP, destination IP, protocol (e.g. TCP or UDP) and IP ID. Thus, it is desirable to ensure that given the same source address, destination address and protocol, the IP ID does not repeat itself in short time intervals.
Simultaneously, the IP ID should not be predictable (across different destination IP addresses) since ``[IP ID] predictability allows traffic
analysis, idle scanning, and even packet injection in specific cases'' \cite{BSD-IPID-new}.

Designing an IP ID generation algorithm that meets both requirements is not straightforward. Since IPv4 was standardized, several schemes have emerged:
\begin{itemize}
    \itemshort Global counter -- This approach  was used in the early IPv4 days due to its simplicity and its non-repetition period of 65536 global packets. However it is extremely predictable and thus insecure, hence abandoned.
    \itemshort Counter/bucket based algorithms -- This family of algorithms, suggested by RFC 7739
    \iflong
    \cite[Section 5.3]{rfc7739},\footnote{While RFC 7739 focuses on IPv6, its proposed algorithms and discussions are also applicable to IPv4.}
    \fi
    is the focus of our work. It uses a table of counters, and a hash function that maps a combination of a source IP address, destination IP address, key and sometimes other elements into an index of an entry in the table. IP ID is generated by choosing the counter pointed to by the hash function, possibly adding  to it an offset (which may depend on the IP endpoints, key, etc.), and finally incrementing the counter. 
    \iflong
    The non-repetition period in this family is 65536 global packets. At the same time, knowing IP ID values for one pair of source and destination IP addresses does not reveal anything about the IP IDs of other endpoints (except those that use the same bucket) -- i.e. it fulfills the non-predictability requirement for almost all other IP destination addresses. This family of algorithms is, therefore, a trade-off between security and functionality.
    \fi
    \itemshort Searchable queue-based algorithm -- This algorithm maintains a queue of the last several thousand IP IDs that were  used. The algorithm draws random IDs  until one is found that is not in the queue. Then this ID is used as the next IP ID, pushed to the queue, and the least-recently used value is popped from the queue. This algorithm ensures  high unpredictability, 
    and guarantees a non-repetition period as long as the queue.
\end{itemize}
Windows (version 8 and later) and Linux/Android implement variants of  the counter-based algorithm.  
MacOS and iOS implement a searchable queue algorithm.

\subsection{Introduction to KASLR}
\iflong
KASLR (Kernel Address Space Layout Randomization \cite{pax-aslr})
\fi
is a security mechanism designed to defeat attack techniques such as ROP (Return-Oriented Programming \cite{Roemer:2012:RPS:2133375.2133377}) that rely on the predictability of kernel code addresses. KASLR-enabled kernels randomize the kernel image load address during boot, so that kernel code addresses become unpredictable. While, e.g. in the Linux x64 kernel, the entropy of the load address is 9 bits, a brute force attack is deemed irrelevant since each failure usually ends in a system freeze (``kernel panic''). A typical KASLR bypass enables the attacker to obtain a kernel address (from which, addresses to useful kernel code gadgets can be calculated as offsets) without de-stabilizing the system.

\subsection{Our Approach}
The IP ID generation mechanisms in Windows and in  Linux (UDP only)  both compute the IP ID as a function of the source IP address, the destination IP address, and a  key $K$ which is generated when the source machine is restarted and is never changed afterwards. We run a cryptanalysis attack which analyzes  the IP ID  values that are sent by a device and extracts the key $K$. This key can then be used to identify the source device, because subsequent attacks will yield the same key value (until the device is restarted). 

In more detail, IP ID generation in both systems maintains a table of counters and uses a hash function to choose which  counter is used  for each connection. 
It seems hard to deploy an attack based on the \textit{value} of the counter, since each IP ID might depend on a different counter. 
Instead, our attack techniques rely on identifying and exploiting {\em collisions} which map two destination IP addresses to the same counter. This enables us to  extract information about the key  that caused the hash values to collide (Linux), or (in Windows) extract information about the offset of the IP ID from the counter. These values  depend on $K$ and therefore enable us to learn $K$ and identify the machine.

Our approach does not rely on an a-priori knowledge of the counter values. Moreover, after we reconstruct $K$, we can reconstruct the current counter values (in full or in part) by sending traffic to specially chosen IP addresses, obtaining their IP ID values and with the knowledge of $K$, work back the counter values that were used to generate them.

\mypar{Linux/Android KASLR bypass}
Support for network namespaces (part of container technology) was introduced in Linux kernel 4.1. With this change, the key $K$ was extended to include 32 bits of a kernel address (the address of the {\tt net} structure for the current namespace). Thus,  reconstructing $K$  also reveals 32 bits of a kernel address, which suffices to reconstruct the full address and be able to  bypass KASLR.\footnote{Through our IP ID attack we were also able to achieve partial KASLR bypass, and a  partial list of loaded drivers, with regards to Windows 10 RedStone 4. This attack was based on an additional initialization bug in  Windows. However, that bug was repaired in  the October 2018 security update and the corresponding KASLR bypass is not effective anymore.}

\mypar{Conclusion}
In general, our work demonstrates that the usage of a non-cryptographic algorithm for the generation of attacker observable values such as IP ID, may be a security vulnerability even if  the values  themselves are not security-sensitive. This is due to an attacker's ability to extract the  key used by the algorithm generating the values, and use this key to track or attack the system.
    
\subsection{Advantages of our Technique}
\label{sec:advantages}
Tracking machines based on the key that is used for generating the IP ID has multiple advantages: 

\textbf{Browser Privacy Mode:} Since our technique exploits the behavior of the IP packet generator, it is not affected if the  browser runs in privacy mode.
 
\textbf{Cross-Browser:} Since our technique exploits the behavior of the IP packet generator, it yields the same device ID regardless of the browser used. It should be noted that browsers (like Tor browser) that relay transport protocols through other servers are not affected by our technique.
 
\textbf{Network change:} Tracking works across different networks since our technique uses bits of $K$ as a device ID, and  $K$ does not depend on the device's IP address or network.
 
\textbf{The ``Golden Image'' Challenge:} Since each device generates its own key $K$ in a random fashion at O/S restart, even  devices with identical software and hardware will most likely have different $K$ values and thus different device IDs.

\textbf{Not easily turned off:} IP ID generation is built into the kernel, and cannot be modified or switched off by the user. Furthermore, the Windows attack can use simple HTTP traffic.
    The Linux/Android attack requires WebRTC which cannot be turned off for mobile Chrome and Firefox. 

\textbf{VPN resistant:} The device ID remains the same when the device uses an IP-layer VPN.

\textbf{Windows shutdown+startup vs. restart:} The  Fast Startup feature of Windows 8 and later,\footnote{\url{https://blogs.msdn.microsoft.com/olivnie/2012/12/14/windows-8-fast-boot}}
which is enabled by default, saves the kernel to disk on shutdown, and reloads it from disk on system startup. Therefore, $K$ is not re-initialized on startup, and keeps its pre-shutdown value. This means that the tracking technique for Windows survives system shutdown+startup. On restart, in contrast, the kernel is initialized from scratch, and a new value for $K$ is generated, i.e. the old device ID is no longer in effect.

\textbf{Scalability:} Our technique can support billions of devices (Windows, Linux, newer Androids), as the device ID is random, and thus ID collisions are only expected due to the birthday paradox. Thus the probability of a single device not to have a unique ID is very low.
\\
\\
It should be noted that in the Linux/Android case, due to the use of 300-400 IP addresses, the need to ``dwell'' on the page for 8-9 seconds, and (in newer Android devices) the excessive attack time, there are use cases in which the technique may be considered invasive and/or inapplicable.

\iflong
\subsection{Additional Contributions}
In addition to the cross-browser \textbf{tracking technique} for Windows and Linux, 
our work introduces multiple additional contributions. 

With respect to Windows,  we also show
\begin{itemize}
    \itemshort The first \textbf{full public documentation} of the IP ID generation algorithm in Windows 8 and later versions, obtained via reverse-engineering of the relevant parts of Windows kernel {\tt tcpip.sys} driver.
    \itemshort \textbf{Cryptanalysis} of said algorithm, resulting in a \textbf{practical technique} to extract 40-45 bits of its key. This analysis is applicable to all Windows 8 and later operating systems.

    \itemshort A scaled down \textbf{demo} implementation of the Windows tracking technique, using only 15 IP addresses (+2 IP addresses for verification) and providing a 40-bit device ID for a field experiment. We provide results from an extensive in-the-wild experiment spanning 75 networks in 18 countries, demonstrating the practicality and applicability of the technique, and also  demonstrating that IP IDs are rarely modified in transit.
\end{itemize}
    
With respect to  Linux/Android, in addition to a  cross-browser \textbf{tracking technique} we also show a   full \textbf{kernel address disclosure} (KASLR bypass), based on revealing a kernel address which is in the {\tt .data} segment of the kernel image.

\fi

We disclosed the vulnerabilities to Microsoft and Linux. Microsoft fixed the issue in Windows April 2019 Security Update 
\iflong
(CVE-2019-0688).\footnote{\url{https://portal.msrc.microsoft.com/en-US/security-guidance/advisory/CVE-2019-0688}}\footnote{The old (vulnerable) logic is still available for Windows 10 versions below 1903 via a registry setting: if during system startup, the registry key {\tt HKLM\textbackslash{}SYSTEM\textbackslash{}CurrentControlSet\textbackslash{}Services\textbackslash{}Tcpip\textbackslash{}Parameters} contains a value named {\tt EnableToeplitzHashForIPID} of type {\tt DWORD} with data {\tt 0x00000001}, then the Toeplitz-based logic is used to generate the IP ID. By default, this value is absent, hence the new (fixed) logic is used. This registry flag is not in effect for Windows 10 versions 1903 and above.}
\fi
Linux fixed the kernel address disclosure (CVE-2019-10639) together with partially addressing the key-based tracking technique (by extending the key to 64 bits) in a patch\footnote{``netns: provide pure entropy for net\_hash\_mix()''  (\url{https://github.com/torvalds/linux/commit/355b98553789b646ed97ad801a619ff898471b92})} applied to Linux kernel versions 5.1-rc4, 5.0.8, 4.19.35, 4.14.112, 4.9.169 and 4.4.179. For 3.18.139 and 3.16.67, Linux applied a patch\footnote{``inet: update the IP ID generation algorithm to higher standards'' (\url{https://git.kernel.org/pub/scm/linux/kernel/git/stable/linux.git/commit/?id=55f0fc7a02de8f12757f4937143d8d5091b2e40b})} we developed, that extends the key to 64 bits. The key-based tracking technique (CVE-2019-10638) is fully addressed in a patch,\footnote{``inet: switch IP ID generator to siphash''  (\url{https://git.kernel.org/pub/scm/linux/kernel/git/torvalds/linux.git/commit/?id=df453700e8d81b1bdafdf684365ee2b9431fb702})} part of kernel versions 5.2-rc1, 5.1.7, 5.0.21, 4.19.48 and 4.14.124.

\section{The Setting}
We assume that 
device tracking is carried out over the web, using an HTML snippet (which can be embedded by a \nth{3} party site/page). 
The snippet forces the browser to send TCP or UDP traffic (one packet per destination IP suffices) to multiple IP addresses under the tracker's control (8-30 addresses for Windows, 300-400 for Linux/Android).  Ideally, such transmission would be rapid. In our experiments, this can be done in few seconds or less.

For the Windows attack, the tracker needs to choose the IP addresses according to some trivial constraints (the Linux IP addresses are not subject to any constraints).
A discussion of the exact constraints and their trade-offs can be found 
\iflong
in the following sections.
\fi
At the server side, the tracker collects the IP ID values sent by the client to each of the IPs, and computes a device ID  consisting  of bits of the key in the device's  kernel data that  is used  to calculate the IP ID.

Additional scenarios (KASLR bypass and internal IP disclosure) for Linux/Android attacks are described in 
\iflong
Appendix~\ref{Appendix:other-attacks}.
\fi
\section{Related Work}
\label{app:usertracking}
Many tracking techniques were suggested in prior research. At large, proposals can be categorized by their passive/active nature. We use the terminology defined in \cite{Wramner}: 
\begin{itemize}
\itemshort 
A {\it fingerprinting} technique measures properties already existing in the browser or operating system,  collecting a combination of data that ideally uniquely identifies the browser/device without  altering its state. 
\itemshort 
A {\it tagging} technique, in contrast, stores data in the browser/device, which uniquely identifies it. Further access to the browser can ``read'' the data 
and identify the device. 
\end{itemize}
As described  in Section~\ref{sec:intro}, fingerprinting techniques  typically  cannot guarantee the uniqueness of the device ID, in particular with respect to corporate machines  cloned from ``golden images''.
Tagging techniques store data on the device, and as such they are more easily monitored and evaded. 
A comprehensive discussion  of tracking methods can be found in Google Chromium's web page ``Technical analysis of client identification mechanisms'' \cite{chromium_zalewski}. 
\iflong
\subsection{Fingerprinting}
There is a major drawback to fingerprinting techniques, which is that typically they cannot guarantee the uniqueness of the device ID. This problem becomes acute when considering organizations wherein desktops and laptops are cloned from ``golden images'', thus making those devices practically indistinguishable for passive techniques. (Furthermore, since fingerprinting techniques are known and understood nowadays, countermeasures are already deployed against some of these techniques.)
For example, font-based fingerprinting, User-Agent header fingerprinting, WebGL (canvas) fingerprinting, browser plugin/extension fingerprinting, and CPU/GPU performance fingerprinting are all methods that cannot distinguish between systems that are based on the exact same hardware and software. Recently, \cite{DBLP:conf/ndss/CaoLW17} improved the accuracy of some of these techniques, and introduced new variants, and \cite{vastel:hal-01652021} measured the longevity of various fingerprinting techniques, and found it limited (and provided suggestions to increase their longevity). However, none of these works addressed the above fundamental shortcoming. Below we discuss the few fingerprinting techniques that do cover the ``golden image'' scenario:\\
\textbf{DNS-based fingerprinting methods:} \cite{Alaca:2016:DFA:2991079.2991091} suggests using the DNS resolver IP address as a fingerprint. However, in an enterprise (or an ISP, or a campus), multitude of clients use the same resolver, and as such, this DNS-based fingerprinting method does not contribute toward distinguishing among these clients.\\
\textbf{Clock skew:} \cite{tcp-clockskew} describes how to remotely measure an endpoint's TCP timestamp clock skew. However, nowadays the risk of enabling TCP timestamps is well understood, and in Windows 7 and later, this feature is disabled by default. \cite{EURECOM+5664} describes how to measure the CPU clock skew using Javascript, however they do not compute a unique ID, but rather attempt to match a signature of a previous measurement, which, even with 300 devices, resulted in at least 20\% multiple matches.\\
\textbf{Using the Javascript Math.random() seed:}
This attack (\cite{math-random-attack}) was addressed by browser vendors in 2008-2010 and is no longer effective.\\
\textbf{Ephemeral source ports in outgoing requests:}
This technique does not work behind a firewall/NAT, as oftentimes the firewall/NAT replaces the original client source port with a port from its own pool. Furthermore, the ports are drawn from a small space (up to $2^{16}$ values) which leads to collisions, and moreover, it requires constant monitoring to track devices.\\
\textbf{Motion sensors (in mobile devices):}
it is possible to fingerprint mobile devices in the browser using deviations in their motion sensors. In general, these techniques are limited in their coverage to such mobile devices that have the required motion sensors. According to \cite{DBLP:journals/corr/BojinovMNB14}, deviations in the accelerometer readouts can be used as a fingerprint, but this requires the mobile device not to move while the measurement takes place. \cite{DBLP:conf/ndss/DasBC16} suggests using the accelerometer and gyroscope, but their calibration process is time consuming, and the accuracy reported (93\%) is insufficient for large scale deployments.\\
\textbf{User action history:} Except for DNS (see below), privacy mode renders this technique ineffective.

\subsection{Tagging}
In general, tagging methods are a well understood privacy threat. Therefore, one of the goals of the privacy modes that are  implemented in major browsers is to make the tagging methods
identify a private browsing session  as  a distinct instance, which has  a different tag than  that of the ``regular'' browser. 
Private browsing sessions  also typically clear their residual data upon termination and start with an empty set of data when launched.

A summary of the privacy mode boundary-crossing status of many tagging techniques appears in~\cite[Table IV]{DBLP:journals/corr/BujlowCSB15}. As is depicted in that table, almost all tagging techniques do not cross the private browsing boundary for most browsers. 
In general, nowadays tagging attempts are blocked by the relevant software vendors. For example, Flash cookies do not cross the privacy mode boundary (\cite{LSO-privacy}), and anyway, Flash nowadays requires user interaction in order to run.

There are some advanced tagging techniques that are not covered in \cite{DBLP:journals/corr/BujlowCSB15}, and yet do not cross the privacy mode boundary: \\
\textbf{TLS-based:} The TLS token binding protocol specifically requires browsers to separate privacy mode tokens from the regular browser tokens 
(\cite[Section 7.3]{ietf-tokbind-protocol-16}). 
Firefox provides a separation between the regular browser and privacy mode with respect to TLS session identifiers and session tickets (\cite{ff-tls}), and likewise Chrome (\cite{chrome-tls}). Recently, a similar technique, using TLS 1.3's session resumption (and session tickets) was described by \cite{TLS-session-resumption}, but it suffers from the above same drawbacks.\\
\textbf{HSTS-based:} HSTS data does not cross the privacy mode boundary in Chrome\cite{chromium_zalewski}, Firefox\cite{HSTS-Safari-Firefox} and Safari\cite{HSTS-Safari-Firefox}.\\
\textbf{HTTPS Public Key Pinning (HPKP):}
\cite{HPKP-HSTS-tagging} describes a tagging technique based on HPKP. However, HPKP is being deprecated - it is only supported nowadays by Firefox.\\
\textbf{DNS cache fingerprinting/tagging (timing based):}
A DNS-based fingerprinting method is proposed in \cite{Felten:2000:TAW:352600.352606}, which can reveal elements of the user's browsing history. This fingerprinting method could in theory be converted to a tagging method. However the ``read tag'' operation is destructive as it changes the data (the tag). \\
\textbf{DNS cache-based tagging:}
Recent work~\cite{KP18} describes a tagging technique based on client side caching of resolved DNS names, where the resolution contains random elements which provide statistic uniqueness. This technique does not work across different networks (as clients typically flush their DNS cache when connecting to a new network), and its longevity is limited by the TTL cap imposed by resolvers and stub resolvers.

\fi

\subsection{IP ID Research}
\label{app:ipidresearch}
\textbf{Device tracking via IP ID:}
Using IP ID is proposed in \cite{Bellovin:2002:TCN:637201.637243} (2002) to detect multiple devices behind a NAT, assuming an IP ID implementation using a {\em global} counter. But nowadays none of the modern operating systems implements IP ID as a global counter. A similar concept is presented by \cite{DBLP:conf/cans/OreviHZ18} for a single destination IP (the DNS resolver) which theoretically works for devices that have per-IP counter (Windows, to some extent). However, this technique does not scale beyond a few dozen devices, due to IP ID collisions (the IP ID field provides at most $2^{16}$ values), and requires ongoing access to the traffic arriving at the DNS resolver.\\
\textbf{Predictable IP ID:} 
The predictability of IP ID may theoretically be used in some conditions to track devices. \cite{Gilad:2013:FCV:2445566.2445568} describes a technique to predict the IP ID of a target,
but requires the adversary to have a fully controlled device alongside it behind the same NAT. 
Also this technique only handles sequential increments (e.g. not time-based). As such, it is inapplicable to the more general scenarios handled in this paper. 
This technique is then used in \cite{DBLP:journals/corr/abs-1205-4011} to poison DNS records.\\
\textbf{OS Fingerprinting:} \cite{SotW} suggests using $\IPID=0$ as a fingerprint for some operating systems.\\
\textbf{Measuring traffic:}  \cite{Shulman:2014:PBP:2665943.2665959} samples IP ID values from servers whose IP ID is a global counter, to estimate their outbound traffic.\\
\textbf{IP ID Algorithm Categorization:} \cite{DR:PAM-18a} provides practical classification of IP ID generation algorithms and measurements in the wild.\\
\textbf{Fragmentation attacks:} While not directly related to the properties of the IP ID field, it should be noted that attack techniques abusing fragmentation are known. RFC 1858 \cite{rfc1858} lists several such attacks, e.g. the ``tiny frgament'' attack and the ``overlapping fragement'' attack.

\textbf{Windows IP ID research:} In parallel to our research,
Ran Menscher published on Twitter his research on Windows IP ID \cite{menscher}. That research  reverse-engineered part of the Windows IP ID generation algorithm (without revealing  how the index to the counter array is calculated). The analysis of this algorithm is based on two assumptions: (1) that the technique is applied shortly after restart, when the relevant memory buffer contains zeroes in a large part of its cells; and (2) that the attacker controls or monitors traffic to pairs of IP addresses which differ in single, specific bit position (including positions in the left half of the address). Based on these extreme assumptions, the attacker can extract the key easily, and use it to expose kernel 31-bit data quantities (though without learning where in the array this data resides).

The uninitialized memory issue exploited by this attack was fixed in Microsoft's October 2018 Security Update \cite{CVE-2018-8493}, which invalidated assumption (1), rendering Menscher's attack completely ineffective. Our attack and our demo, on the other hand, still work against systems that were patched with this update.
Our work has multiple contributions over Menscher's attack:
(1) We provide the full details of the IP ID algorithm. (2) Our analysis does not rely on  the array data, and is thus still in effect after applying the October 2018 Security Update which initializes the array with random data. (3) Our analysis  does not require the  extreme requirements on the relations between the addresses of the controlled/monitored IP addresses. (4) Our  kernel data exposure provides positions of the data, not just data quantities (though our kernel exposure technique, too, was eliminated with the October 2018 Security Update). (5) It should also be noted that unlike our attack,  Menscher's technique could not be used for tracking, since as the cell arrays become  non-zero when they are incremented, the attack becomes ineffective.

\subsection{PRNG seed/key extraction}
Our approach involves breaking the random number generator algorithm used by operating systems to generate the IP ID value and obtain the seed/key used by the algorithm. Similar strategies were used to different ends. For example, \cite{Kumar:2005:EUS:1251086.1251119} broke the PRNG of the Witty worm to obtain the seed, from which they learned the infection time of the Internet nodes. \cite{math-random-attack} broke the Javascript {\tt Math.Random()} PRNG of several browsers, obtained the seed and used it as a browser instance tracking ID. \cite{patent:9384034} broke the {\tt Math.Random()} PRNG of Adobe Flash, obtained the seed and used it to extract the machine clock speed.

\section{Tracking Windows 8 (and Later) Devices}
In this section we first present the algorithm that is used for generating the IP ID in Windows 8 (and later) devices. The input to this algorithm includes a key which is generated at system restart.  
We then  describe how a remote server can identify 45 bits of this key. This data enables to remotely and uniquely identify machines. 
\subsection{IP ID Generation}
\label{Sec:Windows-IPID-Gen}
\paragraph{IP ID prior to Windows 8}
In versions of Microsoft Windows up to and including Windows 7, the IP ID was generated sequentially and globally. That is, for each outgoing IP packet, a global counter would be incremented by 1 and the result (truncated to 16 bits) would be used \cite{DBLP:conf/cans/OreviHZ18}. These older Windows versions are out of scope for this paper.

The source code of the algorithm that is used for generating IP ID values in Windows is not public. However, we recovered the exact algorithm using reverse engineering, and verified its correctness by comparing its output to IP ID values generated by live Windows systems.

\mypar{Technical details}
The algorithm was obtained by reverse-engineering parts of the {\tt tcpip.sys} driver of 64-bit Windows 10 RedStone 4 (April 2018 Update, Build 1803). Apparently this algorithm is in use starting with Windows 8 and Windows Server 2012. 
\iflong
It was positively tested with Windows 8.1 (64-bit), Windows 10 (64-bit), Windows 2012, Windows 2012 R2 and Windows 2016. The algorithm was verified with TCP and UDP over IPv4.
The 32 trailing zero bits used in the calculation of the $v$ variable are hard-wired in the code.
\fi
Notice that the code is not specific to IPv4, and can be used with IPv6, which is why the key $K$ is defined as 320 bits - more than required to support IPv4.\footnote{
Our tracking technique can be probably adapted to IPv6, but since IPv6 is out of scope for this paper, we did not test this.
} For IPv4 pre RedStone 5, only 106 key bits   
\iflong
--- $K_{17,\ldots,94}$ (78 bits), $K1_{17,\ldots,31}$ (15 bits) and $K2_{19,\ldots,31}$ (13 bits) ---
\fi
are used.

\iflong
Henceforth, the term ``IP'' is used as a synonym to ``IPv4''. Also, in order to simplify the discussion, it is assumed that $v$ is returned modulo $2^{15}$ even with Windows 10 RedStone 5 (October 2018 Update, Build 1809), i.e. the most significant bit of the IP ID is simply discarded in this case. 
\fi

\mypar{Toeplitz hash}
The IP ID generation  is based on the Toeplitz hash function defined in \cite{Toeplitz}. 
Let us first define the \textit{Toeplitz hash}, $T(K,I)$, which  is a bilinear transformation from a  binary vector $K$ in  $GF(2)^{320}$, and an input which is a binary string $I$  (where $|I| \leq 289$) to the output space $GF(2)^{32}$. For a binary vector $V$, denote by $V_{i}$ the $i$-th bit in the vector, with bit numbering starting from 0. 
The $i$-th bit  of $T(K,I)$ ($0 \leq i \leq 31$) is defined as the inner product between $I$ and a substring  of $K$ starting in location $i$. Namely
\begin{equation} \label{eq:T-def}
T(K,I)_{i}=\bigoplus_{j=0}^{|I|-1}I_{j} \cdot K_{i+j}
\end{equation}

\mypar{IP ID generation}
The IP ID generation algorithm itself uses keys $K$ ({\tt tcpip!TcpToeplitzHashKey}) which is a 320 bit vector, and $K1$ and $K2$ which are 32 bits each. 
All these keys are generated once during Windows kernel initialization (using {\tt SystemPrng} and {\tt BCryptGenRandom}).

In addition to these constant keys, the algorithm uses  a dynamic array of $M$   counters, denoted $\beta[0],\ldots,\beta[M-1]$, where $M$ is a power of 2, and is specifically set to $M=8192$. 

Algorithm \ref{alg-windows-ipid} describes how Windows 8 (and later) generates an IP ID for a packet delivered from $\IP_{SRC}$ to $\IP_{DST}$, while updating a counter in $\beta$. 

The algorithm uses the keys, and the source and destination IP addresses, to 
pick a random index $i$ for a counter
in $\beta$, and  an offset. The algorithm outputs the sum of the counter $\beta[i]$ and the offset, and increments the counter.  
\begin{algorithm*}
\caption{Windows 8 (and later) IP ID Generation}
\label{alg-windows-ipid}
\begin{algorithmic}[1]
\Procedure{Generate-IPID}{}
\State $i \gets \Num(K2 \oplus T(K,(\IP_{DST})_{0,\ldots,\frac{|\IP_{DST}|}{2}-1}) \oplus T(K,\IP_{SRC})) \mod M$ 
\State $v \gets \beta[i]+\Num(K1 \oplus T(K,\IP_{DST}||\IP_{SRC}||0^{32})) \mod 2^{32}$  
\State $\beta[i] \gets (\beta[i]+1) \mod 2^{32}$
\State return $v \mod 2^{15}$ \Comment{$v \mod 2^{16}$ for Windows 10 RedStone 5}
\EndProcedure
\end{algorithmic}
\end{algorithm*}

\mypar{Notation}
We use  the notation 
$\Num(a_0,a_1,\ldots,a_{31})$ for the number represented in binary by the bits $a_i$, namely the number $\sum_{i=0}^{31} a_i \cdot 2^{31-i}$. (Network byte order is used throughout the paper for representing IP addresses as bit vectors, e.g. 127.0.0.1 is 01111111.00000000.00000000.00000001.)
\iflong
For a vector $V=(V_0,\ldots,V_n)$, denote by $V_{i,\ldots,j}$ the sub-vector $(V_i,\ldots,V_j)$.
\fi

\mypar{Properties of the Toeplitz hash}
\label{Toeplitz-properties}
Our attack uses  the following properties of $T$, which follow from the linearity of this transformation:
\begin{equation} \label{eq:T-zero-trail}
    T(K,I||(0,0,\ldots,0))=T(K,I)
\end{equation}
Therefore the  trailing zeros in the input of $T$
in  the computation of $v$
on line 3 of Algorithm \ref{alg-windows-ipid},  have no effect on the output.
Also,  
\begin{equation} \label{eq:T-contact}
    T(K,I_1||I_2)=T(K,I_1) \oplus T(K,0^{|I_1|}||I_2)
\end{equation}
Therefore it is possible to decompose the second input of  $T$ to two parts, and rephrase the computation as the XOR of two separate expressions. 

\subsection{Reconstructing the Key {\em K}}
\label{sec:reconstructingkeywindows}
To reconstruct the key, the device needs to be measured. The measurements only take a few seconds, and are thus assumed to take place from the same network. I.e., the {\em device's} source IP address, $\IP_{SRC}$, is fixed (though possibly unknown).
A first set of measurements directs the client device to $J$ IP addresses from the same class B network. 
A second set of measurements directs the client device to $G$ \textit{pairs} of IP addresses, each pair in the same class B network, with $G$ different class B network pairs in the set.

Once the device is measured, the attack proceeds in two phases.
The first phase of the attack recovers 30 bits of the key using the first set of measurements.
The second phase of the attack reveals additional 15 bits of the key using 
\iflong
the 30 bits recovered in the first phase and 
\fi
the second set of measurements.
Overall, the measurements reveal 45 bits of the key, which suffices to uniquely identify machines from a large population, with high probability. 

Section~\ref{Appendix:Windows-choosing-G-and-J} describes how to optimally choose the parameters $J$ and $G$ given limits on the number of IP addresses that are available ($L$) and the processing time that is allowed ($\mathcal{T}$). For $L=30$ IP addresses (typical low budget limit), and attack run time limit of $\mathcal{T}=1$ seconds on a single Azure B1s machine ($\alpha=0.001$ from Section~\ref{Windows-results}), the optimal parameter values are $J=6, G=12$.

\subsection{Extracting Bits of {\em K} - Phase 1}
\label{Sec:Windows-phase1}
Denote by $\IP^{g,j}$, $\IPID^{g,j}$ and $\beta[i_g]^{g,j}$ the values of the destination IP address, the IP ID  and $\beta[i]$ (prior to increment) respectively, with respect to the $j$-th packet in the $g$-th class B network that is used in the attack ($j$ and $g$ are counted 0-based).
The first phase of the attack uses only a single class B network, and therefore $g$ is set to 0 in this phase. 
We thus use the following shorthand notation: $\IP^j=\IP^{0,j}$, $\IPID^j=\IPID^{0,j}$ and $\beta_g=\beta[i_g]^{g,0}$.

A major observation is that
only the first half of $\IP_{DST}$ is used to calculate $i$ in Algorithm~\ref{alg-windows-ipid}.
Therefore   packets that are sent to different IP addresses in the same class B network, have an identical   index $i$ into the counter table, and use the same counter $\beta[i]$.  Denote the value of $i$ for the $g$-th class B network as $i_g$.

If these packets are sent in rapid succession (i.e. when no other packet is sent in-between with $i=i_g$), then $\beta[i_g]^{g,j} = \beta_g+j \mod 2^{32}$, and therefore the output in line 5 of the algorithm is calculated with  $\beta[i_g]^{g,j} 
= \beta_g+j \mod 2^{15}$ (for simplicity, in Windows 10 RedStone 5, we discard the most significant bit of the IP ID).

We focus in this phase on the first class B network, $b_0$, with $J$ destination IP addresses in it. 
Note that the offset that is  calculated in line 3 is the difference between the IPID and the counter $\beta[i_0]$ prior to its increment. 

The attack enumerates over the values of the $\beta_0 \mod 2^{15}$ 
\iflong
counter.\footnote{Actually, we show in Appendix \ref{app:incorrect-beta} that one bit of this value is canceled out by the algorithm, and therefore the attack enumerates over  only 14 bits. We ignore this fact for now to simplify the exposition.} 
\fi
For each possible value it calculates the differences between the observed $J$ IPIDs and the corresponding values of the counter, arriving at the  offsets calculated in line 3. By observing pairs of IPIDs, it is possible to identify the correct value of   $\beta_0 \mod 2^{15}$ as well as 30 bits of the key. 

In more detail, for each possible value of $\beta_0 \mod 2^{15}$ the attack calculates the difference 
$$\IPID^j-(\beta_0+j \mod 2^{15}) \mod 2^{15}$$
which, for the right value of the counter should be equal to the offset that is calculated  in line 3. Namely to
$$\Num(K1 \oplus T(K,\IP^{j}||\IP_{SRC}||0^{32})) \mod 2^{15}$$ 
This  value can be expressed as   $(K1 \oplus T(K,\IP^j||\IP_{SRC}||0^{32}))_{17,\ldots,31}$. 
Applying eq.~(\ref{eq:T-zero-trail}) and eq.~(\ref{eq:T-contact}), this expression is simplified into:
$$(K1 \oplus T(K,\IP^j) \oplus T(K,0^{32}||\IP_{SRC}))_{17,\ldots,31}$$.

The attack takes two different $j$ values and computes the XOR of the  
two corresponding  such quantities. This results in the following expression (where we 
denote by $\Vectorize$ a representation of a number in $[0,2^{32})$ as a vector in $GF(2)^{32}$):
$$(\Vectorize(\IPID^j-(\beta_0+j) \mod 2^{15}) \oplus $$
$$\Vectorize(\IPID^{j'}-(\beta_0+j') \mod 2^{15}))_{17,\ldots,31}=$$ 
$$T(K,\IP^j \oplus \IP^{j'})_{17,\ldots,31}$$

This yields  15 linear equations ($i=17,\ldots,31$) on $K$ since (from eq.~(\ref{eq:T-def})):
$$T(K,\IP^j \oplus \IP^{j'})_i=\bigoplus_{m=0}^{31}(\IP^j \oplus \IP^{j'})_{m} \cdot K_{i+m}$$
Since all $\IP^j$ belong to the same class B network, $\IP^j \oplus \IP^{j'}$ always has  0 for its first 16 bits, and therefore $m$ can start at 16. 
Due to obvious linear dependencies, only $J-1$ sets of such equations are useful (e.g. all pairs with $j'=0$), with a total of $15(J-1)$ linear equations for bits $K_{33},\ldots,K_{62}$. That is, for $j=1,\ldots,J-1$ and $i=17,\ldots,31$, the equations are:
\begin{multline} \label{eq:phase1-main}
    \bigoplus_{m=16}^{31}(\IP^j \oplus \IP^0)_{m} \cdot K_{i+m}=\\
    (\Vectorize(\IPID^j-(\beta_0+j) \mod 2^{15})\; \oplus \\
    \Vectorize(\IPID^0-(\beta_0) \mod 2^{15}))_i
\end{multline}

\mypar{Speeding up the computation using preprocessing}
The coefficients of $K$ in eq.~(\ref{eq:phase1-main}) are controlled by the server and are known at setup time. Therefore it is possible to preprocess the computation  of Gaussian elimination. Namely, compute a matrix $Z$ that, when  multiplied by the observed values, reveals  bits of the key.  
\iflong
This preprocessing is important for efficiency, 
and we describe in Appendix~\ref{app:windowspreprocessing} how it can be done. Readers who are only interested in the feasibility of the attack and not in its details can skip that appendix. 
\fi

\paragraph{Attack summary} 
\begin{enumerate}
\itemshort The tracker needs to control $J$ IP addresses in the same class B network.

\itemshort During setup time, the tracker calculates, using  Gaussian elimination, a matrix $Z \in GF(2)^{15(J-1)  \times 15(J-1) }$, based on the values of these IP addresses. 

\itemshort In real time, the tracker gets IP ID values from the device, from packets sent to the $J$ destination IP addresses under the tracker's control. 

\itemshort The tracker then guesses 14 bits ($\beta_0 \mod 2^{14}$ - the most significant bit of $\beta_0 \mod 2^{15}$ cancels itself in eq.~(\ref{eq:phase1-main})) of the counter that is used for these IP addresses, calculates vectors $D^j$ ($j=1, \ldots, J-1$), 
\iflong
that are defined as differences of functions of the observed IP IDs (details in Appendix~\ref{app:windowspreprocessing}) 
\fi
and performs a matrix-by-vector multiplication of $Z$ and the vector $(D^0,\ldots,D^{J-1})$. 

For the correct value of $\beta_0 \mod 2^{14}$ 
this computation results in a  vector of $15(J-1)$ bits, whose first 30 bits are $K_{33},\ldots,K_{62}$ and the remaining bits are zero. 
\itemshort The attacker identifies the right value of the counter by comparing  to zero the   
$15(J-1)-30$ bits starting at position 31: if $15(J-1)-30 \gg 14$, this verification statistically guarantees the correctness of the solution (up to a flipped most significant bit in $\beta_0 \mod 2^{14}$, 
\iflong
see Appendix \ref{app:incorrect-beta}.)
\fi
\end{enumerate}

Overall this process reveals 30 bits of the key as well as the value  $(\beta_0 \mod 2^{14})$. 

The attack takes  $2^{14}\cdot (15(J-1))^2$ bit operations  (for enumeration over the possible key values and for the  matrix-by-vector 
\iflong
multiplication),\footnote{Run time  can be improved by  conducting the comparison to zero on a vector-by-vector basis, eliminating on the first non-zero value encountered. This requires  an average of $2 \cdot 2^{14}(15(J-1))$ bit operations} 
\fi
and $(15(J-1))^2$ memory bits (for $Z$).  
As explained in  Section~\ref{Appendix:Windows-choosing-G-and-J}, we set $J=6$ and therefore this overhead is very small. 

The tracker  obtains the (correct) value $\beta_0 \mod 2^{14}$, which will be used in the next phase. While it is guaranteed that the correct $K$ and $\beta_0 \mod 2^{14}$ will be found, the algorithm may emit additional candidates (with incorrect $\beta_0 \mod 2^{14}$). The false positive probability of both phases of the attack is analyzed in 
\iflong
Appendix \ref{appendix:false-positives}.\footnote{Note: Throughout the paper, we assume that $\rank(C)=30$. This results in a single key vector per guessed $\beta_0 \mod 2^{14}$. We discuss the 
conditions on $\IP^0,\ldots,\IP^{J-1}$ to meet this assumption in Appendix \ref{appendix:rank-ker-C}.
If $\rank(\ker(C))>0$, then each guess of $\beta_0 \mod 2^{14}$ yields $2^{\rank(\ker(C))}$ possible keys.
Thus small values of $\rank(\ker(C))$ are acceptable.}
\fi

\subsection{Extracting Bits of {\em K} - Phase 2}
\label{Section:Phase-2}
Given 30 bits of $K$ ($K_{33},\ldots,K_{62}$) and the value $(\beta_0 \mod 2^{14})$, recovered in Phase 1, the attack can be extended to learn a total of up to 45 key bits ($K_{18},\ldots,K_{62}$). This is done in the following way. 
The offset for $\IPID^0$ computed in line 3 of Algorithm \ref{alg-windows-ipid} is: 
$$\Num(K1 \oplus T(K,\IP^0) \oplus T(K,0^{32}||\mathit{\IP}_{SRC})) \mod 2^{15}=$$
$$(\IPID^0-\beta_0) \mod 2^{15}$$
The  following equation  follows from the previous one:
$$(K1 \oplus T(K,0^{32}||\IP_{SRC}))_{17,\ldots,31}=T(K,\IP^0)_{17,\ldots,31} \oplus $$
$$\Vectorize(\IPID^0-\beta_0 \mod 2^{15})_{17,\ldots,31}$$

The tracker looks at pairs of IP addresses in the remaining B classes ($b_1,\ldots,b_G$), each pair in a different class B network. Denote each such pair as $(\IP^{g,0},\IP^{g,1})$, with the order inside the pair conforming to the order of packet transmission, and the packets being transmitted in rapid succession. 
Substituting the above into the definition of $\IPID$ yields: 
\vspace{-0.25cm}
\begin{eqnarray*}
\IPID^{g,j} \hspace{-0.4cm} & & =  \beta_g+j+\Num( \; T(K,\IP^0)_{17,\ldots,31} \\
 & &  \oplus \; \Vectorize(\IPID^0-\beta_0 \mod 2^{15})_{17,\ldots,31}  \\
 & &  \oplus \;  T(K,\IP^{g,j})_{17,\ldots,31}\; ) \mod 2^{15} 
\end{eqnarray*}

Using the linearity of $T$, this is simplified into:
\vspace{-0.25cm}
\iflong
\begin{multline}
\label{eq:ipid}
\IPID^{g,j}=\beta_g+j+
\Num\ (\; T(K,\IP^0 \oplus \IP^{g,j})_{17,\ldots,31} \; \oplus  \\
\;\;\;\;\;\; \Vectorize(\IPID^0-\beta_0 \mod 2^{15})_{17,\ldots,31}\; ) \mod 2^{15}
\end{multline}
\fi

\vspace{-0.25cm}
Let us use the notation
\begin{multline*}
S^{g,j}=
\Num\ (\; T(K,\IP^0 \oplus \IP^{g,j})_{17,\ldots,31} \; \oplus  \\
\;\;\;\;\;\; \Vectorize(\IPID^0-\beta_0 \mod 2^{15})_{17,\ldots,31}\; ) \mod 2^{15}
\end{multline*}
\iflong
Then eq.~(\ref{eq:ipid})  becomes
\fi
$$
\IPID^{g,j}=\beta_g+j+ S^{g,j} \mod 2^{15}
$$

Subtracting the IPIDs of the two consecutive packets in the same B class (with $j=0$ and $j=1$) cancels  the value of the counter $\beta_g$, and yields:
\vspace{-0.25cm}
\begin{equation}
\label{eq:phase2-main}
(\IPID^{g,1}-\IPID^{g,0}) \mod 2^{15}= 1 + S^{g,1} - S^{g,0} \mod 2^{15}
\end{equation}

The left side of the equation is observed by the tracker. The right side can be computed based on  $\beta_0 \mod 2^{15}$ and 
\iflong
$K_{17},\ldots,K_{62}$.\footnote{As explained in App.~\ref{app:windowsphase2}, the bit $K_{17}$ does not affect the computation. The same holds for the most significant bit of $\beta_0 \mod 2^{15}$, so knowing $\beta_0 \mod 2^{14}$ suffices for the attack.}
\fi
The tracker already knows these values except for  $K_{18},\ldots,K_{33}$, 
and therefore only needs to enumerate over the $2^{15}$ possible values of $K_{18},\ldots,K_{32}$ and  eliminate all values which do not agree with the equation. We discuss this procedure in  depth in 
\iflong
Appendix~\ref{app:windowsphase2}.
\fi

\paragraph{Attack summary:} 
\begin{enumerate}
\itemshort  The tracker needs to control additional $G$ pairs of IPs (each pair in its own class B network). 

\itemshort Given IP IDs for these pairs, the tracker enumerates over  additional 15 key bits, and then, for each pair of IP addresses, calculates both sides of eq.~(\ref{eq:phase2-main}) and compares them. For this calculation the tracker can choose $K_{17}$ and the leftmost bit of $\beta_0 \mod 2^{15}$ arbitrarily, as they will both cancel themselves. 

\itemshort In theory, each IP pair should yield a $2^{15}$ elimination power for identifying the right key, but see 
\iflong
Appendix \ref{appendix:false-positives} 
\fi
for a more accurate analysis. 

\itemshort In the calculation, the leading term (in terms of run time) is computing $T(K,I)_{17,\ldots,31}$ (where $|I|=32)$, which takes $14|I|$ bit operations, and is used twice. Thus, the run-time is roughly $2^{15} \cdot 2 \cdot 14 \cdot 32$ bit operations (there is no multiplication by $G$ since the first pair is likely to eliminate almost all false guesses). 
\end{enumerate}
At the end of Phase 2, the tracker obtains:
\begin{itemize}
    \itemshort A partial key vector (or some candidates) $K_{18},\ldots,K_{62}$ (45 bits), which is specific to the device since it was set during kernel initialization, and does not depend on $\IP_{SRC}$. These bits 
    serve as a device ID.
    
    \itemshort The value 
    \vspace{-0.3cm}
    \begin{multline*}
    (K1 \oplus T(K,0^{32}||\IP_{SRC}))_{18,\ldots,31}=
    T(K,IP^0)_{18,\ldots,31} \\
    \oplus \Vectorize(\IPID^0-\beta_0 \mod 2^{14})_{18,\ldots,31}
    \end{multline*}
This value allows the tracker to calculate (assuming $K_{18},\ldots\,K_{62}$ are known) the value of the counter
$\beta[i] \mod 2^{14}$
for any destination IP address whose IP ID is known (provided the source IP is $\IP_{SRC}$).\footnote{This is useful for reconstructing the table $\beta$ of counters 
\iflong
(Appendix \ref{appendix:exposing-kernel-data}.)
\fi
 }
\end{itemize} 

\subsection{Choosing Optimal $G$ and $J$}
\label{Appendix:Windows-choosing-G-and-J}
For Windows, we assume budget-oriented constraints, namely $L$ available IP addresses and $\mathcal{T}$ CPU time per measurement. We need  to set the number $J$ of IP addresses from the same class B network to which the client is directed in the  first set of measurements, and the number $G$ of pairs of IP addresses, each pair in the same class B network, used in the  second set of measurements. 

Our goal is to optimize for minimum false positives. The first constraint can be expressed as  $J+2G \leq L$. As for the second constraint, the  leading term of the time of the attack run is $\alpha \cdot (J!)$ (Appendix~\ref{Appendix:Windows-packet-order}), where $\alpha$ expresses the computing platform's strength. Therefore, we can approximate the second constraint as $\alpha \cdot (J!) \leq \mathcal{T}$.
Additionally, there are inherent constraints: $J-1 \geq 3$ to let Phase 1 suggest a single key candidate to Phase 2 (most of the time), and $G \geq 2$ to let Phase 2 provide a single final key (most of the time).  

Given these constraints, we want to minimize the leading term in false positives, $2 \cdot 2^{-\frac{G+J-1}{2}}$ 
\iflong
(Appendix~\ref{appendix:false-positives}),
\fi
i.e. we need to maximize $G+J$. Since we ``pay'' two IP addresses for each increment of $G$ and only one IP address for each increment of $J$, we should make $J$ as large as possible (as long as $G$ is valid), so the solution is:
$$J=\min(\max(\{J\; |\; \alpha J!\leq \mathcal{T}\}),L-4)$$

(As stated in Section~\ref{sec:reconstructingkeywindows}, for $L=30$, $\mathcal{T}=1$ sec.,  and $\alpha=0.001$, the optimal combination is $J=6, G=12$.)

\subsection{Practical Considerations}
\label{Sec:Windows-practical-considerations}
We discuss in Appendix~\ref{app:practical} different issues that appear when deploying the attack. These issues include ways to emit the needed traffic from the browser, handling packet loss and out-of-order packet transmission, handling interfering packets, and limiting the false-positive and false-negative error probabilities. 

The run time of the key extraction attack is less than a second even on a very modest machine. The {\em dwell time} (time duration in which the page needs to be loaded in the browser) is 1-2 seconds for a WebSocket implementation. It is possible to minimize the dwell time by moving to WebRTC (STUN). 

Longevity: the device ID is valid until the machine restarts (mere shutdown+start does not invalidate the device ID due to Windows' Fast Start feature). A typical user needs to restart his/her Windows machine only for some Windows updates, i.e. with a frequency of less than once per month.

The attack is scalable: with 41 bits, the probability of a device to have a unique ID is very high, even for a billion device population; false positives are also rare ($2.1 \times 10^{-6}$ -- Table~\ref{Table:common-tail}), and false negatives can be made negligible (Appendix~\ref{App:Windows-false-negatives}).
From resource perspective, the attack uses a fixed number of servers, RAM/disk and ($L=30$) IPs. The required CPU power is linear in the number of devices measured per time unit, and in the Windows case is negligible. Network consumption per test is also negligible 
\iflong
(assuming WebRTC/STUN implementation, a single STUN binding request is 48 IP level bytes, thus the total network traffic consumed is $48L$ IP layer bytes in $L=30$ packets, i.e. less than 1.5KB at the IP layer.)
\fi

\subsection{Attack Improvements and Variants}
\label{Sec:Windows-attack-improvements}
A {\bf fast-track identification of already-seen keys} can be obtained in the following way:  
Once bits of a key $K$ are extracted, they will be  stored for comparison against future connections.  When a device is to be measured, the tracker first goes through all stored $K$ bit strings, and  tests the measured data for compatibility with each one of them. This amounts to guessing the bits of $\beta_0$ one by one, starting from the least significant, and eliminating via eq.~(\ref{eq:phase2-main}), using $\mod 2^{n}$ where $n$ is the number of $\beta_0$ bits guessed so far. The CPU work per key is thus almost negligible.

The original attack can also be sped up using {\bf incremental evaluation}. The details are in
\iflong
Appendix~\ref{sec:incrementalenumeration}.
\fi

\subsection{Environment Factors}
\label{sec:env-factors}
We demonstrate here that the tracking attack can be deployed in almost every setting that can be reasonably expected.  

\textbf{HTTPS:} In essence, there should be no problem in having the snippet use WebSocket over HTTPS ({\tt wss://} URL scheme) for TCP packets.

\textbf{NAT:} Typically NAT (Network Address Translator) devices do not alter IP IDs, and thus do not affect the attack.

\textbf{Transparent HTTP Proxy / Web Gateway:} Such devices may terminate the TCP connection and establish their own connections (with IP ID from their own network stack) and thus render our technique completely ineffective. However, typically these devices do not interfere with HTTPS (TCP port 443) traffic, and UDP traffic, so these alternatives can be used by the tracker.

\textbf{Forward HTTP proxy:} When a browser is configured to use a forward proxy server, even HTTPS traffic is routed to it by the browser. However, it may still be the case that UDP traffic (which is not handled by HTTP forward proxies) can be used by the technique.

\textbf{Tor-based browsers and similar browsers:} Browsers that forward TCP traffic to proxy servers (and disallow or forward UDP requests) are incompatible with the tracking technique as they do not expose IP header data generated on the device. Since ``Tor transports TCP streams, not IP packets'',\footnote{\url{https://www.torproject.org/docs/faq.html.en\#RemotePhysicalDeviceFingerprinting}}
this applies to all Tor-based products, such as the Tor browser and Brave's ``Private Tabs with Tor'' and therefore they are not covered by our technique.

\textbf{Windows Defender Application Guard (WDAG):} This new technology in Windows 10 enables the user to launch the Edge browser in a virtual environment. While the device ID in this virtual environment is independent of the device ID of the main operating system, it is consistent among {\bf all} WDAG Edge instances. Furthermore, unlike the ``main'' Windows device ID, the WDAG device ID does not change with operating system restart, hence the WDAG device ID lives longer than the main Windows device ID. It should be noted that WDAG is only available for Edge browser in Windows 10 Enterprise/Pro edition, and requires high-end hardware.

\textbf{IP-Level VPN:} We experimented with F-Secure FreeDome (\url{www.f-secure.com/en/web/home_global/freedome}) 
and PureVPN (\url{www.purevpn.com/}). 
Both VPNs supported our technique.

\textbf{IPv6 and IPsec:} We do not know whether IPv6 or IPsec packets use the same IP ID generation mechanism. This requires further research.

\textbf{Javascript disabled:} Tracking can also work when Javascript (or any client side scripting) is not available, e.g. with the NoScript browser extension \cite{noscript}. We discuss this in 
\iflong
Appendix~\ref{app:nojavascript}.
\fi

\subsection{Possible Countermeasures}
\label{Windows-mitigation}
We list here some obvious ways of modifying Algorithm \ref{alg-windows-ipid} and their impact:
\begin{itemize}
    \itemshort Increasing $M$ (the size of the table of counters) -- surprisingly, this has very little effect on the basic tracking technique, since no assumptions were made on $M$ in the first place. It does affect the $\beta$ reconstruction technique.
    
    \itemshort Changing $T$ into a cryptographically strong keyed-hash function -- while this change eliminates the original attack, it is still possible to mount a weaker attack that only tracks a device while its $\IP_{SRC}$ does not change. In fact, this applies to the entire abstract scheme proposed in \cite[Section 5.3]{rfc7739}. 
    \iflong
    See Appendix \ref{appendix:attacking-crypto-T}.
    \fi
    
    \itemshort Changing the algorithm altogether (this is our recommendation). A robust algorithm relies on industrial-strength cryptography, large enough key space, and strong entropy source for the key, and uses them to generate IP IDs which (a) have guaranteed non-repetition period; (b) are difficult to predict; and (c) do not leak useful data. The algorithm used in macOS/iOS \cite{BSD-IPID-new} is a good example. This eliminates the attack altogether.
\end{itemize}

\section{Field Experiment -- Attacking Windows Machines in the Wild}
\label{Sec:windows-experiment}
We set up a fully operational system to test the IP ID behavior in the wild, as well as to verify that the  technique for extracting device IDs for Windows machine works as expected.
\subsection{Setup}
\label{Sec:Windows-setup}
As explained in
\iflong
Appendices \ref{appendix:handling-false-positives} and \ref{appendix:false-positives},
\fi
in order to avoid false positives (which almost always happen due to false keys that differ from the true key in a few most significant bits), we need to trim the most significant bits from the key -- i.e. use the key's tail. For the full production setup (30 IP addresses), we calculated that a tail of 41 bits will suffice. Due to logistic and budgetary constraints, in our experiment we used only 15 IP addresses (rather than 30) for the key extraction (and 2 more IPs for verification), with $J=5, G=5, Q=1$. Thus we lowered the tail length to 40, and used the 40 bits $K_{23},\ldots,K_{62}$ as a device ID. That is, for this experiment, we traded the device ID space size for a smaller probability of false positives.

We then used WebSocket traffic to the additional pair of IP addresses (from a  class B network that is different than those in the initial set of 15 IPs) to verify the correctness of the key bits extracted. In this experiment, since we do not extract $K_{17,\ldots,22}$ we can only compute the least significant 9 bits of the IPID, adapting eq.~(\ref{eq:phase2-main}) into:
$$\IPID^{g,1} \mod 2^9= \IPID^{g,0} \pm 1 + S^{g,1} - S^{g,0} \mod 2^9$$
(We need to use $\pm 1$ since we cannot know the order of packet generation. Thus 
given knowledge of $\IPID^{g,0}$ 
we have two candidates for $\IPID^{g,1}$, out of a space of $2^9=512$ values.) A random choice of two values yields a success rate of $\nicefrac{1}{256}$. We deem our algorithm to be valid if it consistently yields the correct value (in one of the candidates) in all tests.

\iflong
Both regular web traffic (e.g. snippet download) and WebSocket traffic were carried out in the clear, over HTTP ports 80 and 8080 respectively.
\fi

We asked ``Friends and Family'' to browse to the demo site using Windows 8 or later, from various networks.
\subsection{Results}
\label{Windows-results}
\paragraph{Network distribution}
The experiment was conducted from July \nth{22}, 2018 to October \nth{20}, 2018. We collected data on 75 different class B 
\iflong
networks.\footnote{Except for 3 tests in the same class B network. In two of these tests, the scenarios were of a user roaming from abroad having the same class B as local cellular network access. In one case the subnet ownership was different.}
\fi
\iflong
The networks are well dispersed across 18 countries and 4 continents with representatives from Australia, Austria, Belgium, Canada, Denmark, Finland, France, Germany, Hong Kong, Israel, India, Japan, The Netherlands, Poland, Sweden, Switzerland, UK and USA.
\fi
The networks are also usage-diverse (home networks, SMB networks, corporate networks, university networks, public hotspots and cellular networks). 
We asked the users who connected to our demo site to use multiple regular browsers
\iflong
(indeed the snippet was accessed with all the common Windows browsers)
\fi
and networks, and connect at different times, and verified that the device ID remained the same in all these connections.  
\mypar{Failures to extract a key -- IP ID modification}
In only 6 networks out of 75 (8\%) we could not extract the key and therefore concluded that the IP ID was not preserved by the network.
These six networks  did not include any major ISP and seem to be used by relatively few users: they included an airport WiFi network, a government office, and a Windows machine connecting through one cellular hotspot (hotspots that we tested in other cellular networks did not change the IP ID).
Of those six networks, in 3 networks we had clear indication
\iflong
(via an HTTP request header - {\tt Via} or {\tt X-BlueCoat-Via})
\fi
that a transparent proxy
\iflong
(in two cases - Squid 3.1.23)
\fi
or a web security gateway
\iflong
(in one case - BlueCoat ProxySG)
\fi
was in path. In such cases, moving to WebSocket over HTTPS, or to UDP would probably have addressed the issue. Another case was a forward proxy (moving to UDP would  have possibly addressed it). 
\iflong
In the two final cases, the exact nature of interference was not identified, however one case clearly exhibited Linux characteristics at the IP and TCP level (hence, it is very likely to be a Linux-based TCP gateway), and the other exhibited non-Windows TCP artifacts (TCP timestamps), thus most likely another TCP gateway. 
\fi
We can say then that optimistically, only 2 networks out of 75 (2.7\%) are incompatible with the tracking technique, maybe even less (as it is still quite possible these two TCP gateways are actually transparent proxies).
\mypar{Positive results}
In the remaining 69 networks, for 4 networks we did not keep traffic for the additional two IPs, thus we could not verify the key extraction. For the rest 65 networks, our algorithm extracted a single 40-bit key, and correctly predicted the least significant 9 bits of the IPID of the second IP in the last pair (i.e. the correct value was one of the two candidates computed by the algorithm). This verifies the correctness of the algorithm and the key bits it extracts.
\mypar{Lab verification}
We tested a machine in the lab with the above test setup to obtain 40 bits of $K$. Then, using WinDbg in local kernel mode, we obtained {\tt tcpip!TcpToeplitzHashKey}, extracted the 40 bits from it and compared to the 40 bits calculated by the snippet -- as expected, they came out identical.
\mypar{Actual run time}
\iflong
Our demo system was implemented on the least powerful (and cheapest) Azure VM (B1s class, 1 vCPU \cite{Azure-VMs}).
Based on the run time measured for Phase 1 with $J=5$ (0.12 seconds), and since the attack in proportional to $J!$, we can compute the CPU speed factor $\alpha$, $\alpha=\frac{0.12}{5!}=0.001$ and using it we estimate the run time for Phase 1 with $J=6$ to be approx. 0.72 seconds. 
We estimate the run time in Phase 2 to be 0.01 seconds or less, thus the overall run time 
would be 0.73 seconds. Extrapolated to 10,000 Azure B1s, the run time would be 0.000073 seconds.
\fi

\mypar{Packet loss and false negatives} 
We analyzed 79 valid 
\iflong
tests\footnote{Some networks were tested multiple times} (with Windows 8+ operating system, and no IP ID modification), 
\fi
and found only 3 cases wherein the analysis logic failed to provide a device ID (additional test from the same devices succeeded in extracting a key). In all such cases a manual analysis indicates that this is due to packet loss. 
\iflong
All three cases were from locations where Internet connectivity is not ideal, and are also geographically remote from our server (the tested networks were in Asia, whereas our server is in the US). In two cases (both belong to the same user, in a cellular network), there were 3 missing packets, and in one case -- 4. \fi
Appendix~\ref{App:Windows-false-negatives} describes additional logic 
\iflong
(not implemented in the experiment)
\fi
that can be used to reduce false negatives to a negligible level.
\section{Linux and Android}
The scope of our research is Linux kernel 3.0 and above. Also, we  only investigated  the x64 (typical desktop Linux) and ARM64 (Android) CPU architectures, although almost all of the analysis is not architecture-specific. 

\subsection{Attack Outline}
\label{sec:linuxoutline}
In order to track a Linux/Android device, the tracker needs to control several hundred IP addresses. The tracking snippet forces the browser to rapidly emit UDP packets to each such IP (using WebRTC and specifically the STUN protocol, which enables sending bursts of packets closely spaced in time to controlled destination addresses). It also collects the device's source IP address (using WebRTC as well 
\iflong
--  \cite{webrtc-attack}
\fi
or 
\iflong
alternatively Appendix~\ref{Appendix:internal-ip-disclosure}.)
\fi

The tracker collects IP IDs from all IP addresses, and identifies  bucket collisions by looking for IP pairs whose IP IDs are in close proximity. Recall that the choice of the bucket is a function of the source and destination IP addresses, and a device key. The tracker enumerates over the key space to find the (correct) key which generates collisions for the same pairs for which  collisions were observed. The key that is found is the device ID.

\subsection{IP ID Generation in  Linux}
\label{sec:linuximplementation1}
The Linux kernel implementation of IP ID differs between TCP and UDP \cite{Knockel2014}.
The TCP implementation always used a counter per TCP connection (initialized with a hash of the connection endpoints and a secret key, combined with a high resolution timer) and as such, is not interesting to us (collisions are meaningless).
The implementation of IP ID for stateless over-IP 
\iflong 
protocols\footnote{This also covers the IP ID of TCP RST packets which do not belong to an established TCP connection, e.g. RST for SYN received to a non-open port, and RST for SYN+ACK received with no matching SYN previously sent.} 
\fi
(e.g. UDP) has gone through an interesting evolution process. We focus on short datagrams, i.e. datagrams shorter than MTU (maximum transmission unit), that do not undergo fragmentation. We designate the IP ID generation algorithms as $A_0, A_1, A_2$ and $A_3$, in their order of evolution.

\textbf{$A_0$:} In early Linux kernels, the IP ID for short datagrams was simply set to 0. 

\textbf{$A_1$ and $A_2$:} In Linux kernel 3.16.0 (released August 2014), IP ID for short datagrams became dynamic (just like it has always been for long UDP datagrams).\footnote{See  function {\tt \_\_ip\_select\_ident} in  \url{https://elixir.bootlin.com/linux/v3.16/source/net/ipv4/route.c}.} This was back-ported to various active Linux 3.x branches (see Table \ref{table:linux-combinations}). The generation algorithm\footnote{Except when the UDP client sets the PMTU discovery mode for the socket to {\tt IP\_PMTUDISC\_DO}, {\tt IP\_PMTUDISC\_PROBE} or {\tt IP\_PMTUDISC\_INTERFACE}. This can be done via {\tt setsockopt(\ldots,IPPROTO\_IP,IP\_MTU\_DISCOVER,\ldots)}. In such case the IP ID algorithm used is fundamentally different, and the attack described in this paper cannot be used with the socket's datagrams. Note that Chrome's gQUIC implementation for Linux and Android sets the PMTU discovery mode to {\tt IP\_PMTUDISC\_DO}, and therefore the attack in this paper does not apply to it.} in general has an array of $M=2048$ buckets, each containing a value $0 \leq \beta < 2^{16}$
\iflong
(the implementation uses 32 bit quantities, but only the least significant 16 bits are used)
\fi
and a time-stamp $\tau$ of the last time this bucket was 
\iflong
used (this time-stamp is taken in a  resolution of $f$ Hz, where $f$ depends on the version of the OS).
\fi
The bucket array is initialized at boot time with random data (using a PRNG).
The algorithm also uses the following parameters
\begin{itemize}
    \itemshort 
$key$ -- a 32-bit key  ({\tt ip\_idents\_hashrnd}) which is initialized upon first IP transmission with random data.
    \itemshort 
$h$ -- a hash function. 
\iflong
The details of the hash function  are not important for  understanding the attack. There are two versions of the hash functions: the old one is used in $A_1$, and the new one is used in $A_2$ and $A_3$.\footnote{In $A_1$ the hash function $h$ is a modified Jenkins lookup3 hash function \cite{Jenkins3}, except that the initialization steps are taken from the Jenkins lookup2 hash function \cite{Jenkins2} but using a different constant ({\tt 0xdeadbeef} instead of {\tt 0x9e3779b9}).
This IP ID algorithm was back-ported to several earlier kernel branches.
In Linux 4.1 (released June 2015), the hash function $h$ was corrected to fully comply with the Jenkins lookup3 hash function. This change was back-ported to Linux 3.18 (ver. 3.18.17) which is the active development 3.x kernel branch, and to several earlier kernel branches. This version is used in $A_2$ and $A_3$.
}
\fi
\itemshort 
 $protocol$ -- the IP ``next level'' protocol number (for UDP, this value is 17). Nominally 8-bit field, extended to 32-bit by zero-filling the most significant bits.
     \itemshort 
$\textsc{random}(x), x>0$ -- a PRNG (a 96/128 bit Tausworthe Generator) which receives $x$ as a parameter and provides a random integer in the range $[0,x)$. (We define $\textsc{random}(0)=0$). Note that $\textsc{random}(1)=0$. 
\end{itemize}

The IP ID generation algorithm is defined in Algorithm~\ref{alg-linux-ipid}. The procedure picks an index to a counter as a function of the source and destination IP address, the protocol and the key. It picks a random value which is smaller than or equal  to the time that passed (measured in ticks, with tick frequency of $f$ per second) since the last usage of this counter, increments the counter by this value, and outputs the result.   
\begin{algorithm}
\caption{Linux IP ID Generation ($A_1$/$A_2$)}
\label{alg-linux-ipid}
\begin{algorithmic}[1]
\Procedure{Generate-IPID}{}
\State $i \gets h(IP_{DST},IP_{SRC},protocol,key) \mod M$ 
\State $hop \gets 1+\Call{random}{t_{now}-\tau[i]}$
\State $\beta[i] \gets (\beta[i]+hop) \mod 2^{16}$
\State $\tau[i] \gets t_{now}$
\State return $\beta[i]$
\EndProcedure
\end{algorithmic}
\end{algorithm}

\textbf{$A_3$:} Starting with Linux 4.1, the net namespace of the kernel context, $net$ (a 64-bit {\em pointer} in kernel space) is included in the hash calculation, conditional on a compilation flag {\tt CONFIG\_NET\_NS} (which is on by default for Linux 4.1 and later, and for Android kernel 4.4 and later). The modification is for step 2, which now reads:
$$i \gets h(IP_{DST},IP_{SRC},protocol \oplus g(net),key) \mod M$$ 
where $g(x)$ is a right-shift (by $\rho$ bits) and a truncation function that returns 32 bits from $x$. We designate this algorithm as $A_3$. 

To summarize, there are four flavors of IP ID generation (for short stateless protocol datagrams) in Linux: 
\begin{enumerate}
\itemshort $A_0$ - IP ID is always 0 (in ancient kernel versions)
\itemshort $A_1\; / \; A_2$ - Both versions use Algorithm \ref{alg-linux-ipid}, with the different implementations of $h$.
\itemshort $A_3$ - 
\iflong
Algorithm \ref{alg-linux-ipid}, with $h$ being a correctly implemented Jenkins lookup3 hash function,  with net namespace.
\fi
\end{enumerate}
Of interest to us are algorithms $A_1$ to $A_3$. We focus mostly on UDP, as this is a stateless protocol which can be emitted by browsers.

The resolution $f$ of the timer $t$ in the algorithm is determined by the kernel compile-time constant {\tt CONFIG\_HZ}. A common value for older Android Linux kernels is 100(Hz). Newer Android Linux kernels (4.4 and above) use 300 or 100 (or rarely, 250).
The default for Linux is $f=250$.\footnote{\url{https://elixir.bootlin.com/linux/v4.19/source/kernel/Kconfig.hz}} In general, for tracking purposes, a lower value of $f$ is better.

Note that $key$ and $net$ are generated during the operating system initialization, which, unlike Windows, happens during restart {\em and} during (shutdown+)start. 

\subsection{Setting the Stage}
\label{sec:linuxsetting}
Our technique for  tracking Android (and Linux) devices
uses HTML5's WebRTC\cite{WebRTC} 
both to discover the internal IP address of the device
\iflong
\cite{webrtc-attack},
\fi
and to send multiple UDP packets. It works best when the WebRTC STUN \cite{rfc5389} traffic is bursty. In order to analyze the effectiveness of the technique  we investigated  the following features, focusing on Android devices.

\label{app:android}
\iflong
\subsubsection{Android Versions and Linux Kernel Versions}
\fi
\label{app:android1}
The Android operating system is based on the Linux kernel. However, Android versions do not map 1:1 to Linux kernel versions. The same Android version may be built with different Linux kernel versions by different vendors, and sometimes by the same vendor.
Moreover, when an Android device updates its Android operating system, typically its Linux kernel remains on the same branch (e.g. 3.18.x). Android vendors also typically use somewhat old Linux kernels. Therefore, many Android devices in the wild still have Linux 3.x kernels, i.e. use algorithm $A_1$ or $A_2$. 

\iflong
\subsubsection{Sending Short UDP Datagrams to Arbitrary Destinations, or ``Set Your Browsers to STUN''}
\fi
\label{sec:STUN}
The technique requires sending UDP datagrams from the browser to multiple destinations. The content of the datagrams is immaterial, as the tracker is interested only in the IP ID field. 
We use WebRTC (specifically -- STUN) to send short UDP datagrams (with no control over their content) to arbitrary hosts. The {\tt RTCPeerConnection} interface can be used to instruct the browser's WebRTC engine to use a list of presumably STUN servers, and even allows setting the UDP destination port per each host.
The browser then sends STUN ``Binding Request'' (UDP short datagram) to the destination host and port. 

To send STUN requests to multiple servers (in Javascript), create an array {\tt A} of strings in the form {\tt stun:{\it host}:{\it port}}, then invoke the constructor {\tt RTCPeerConnection(\{iceServers: A\}, \ldots)} in a regular WebRTC flow e.g. \cite{RTCDataChannel-for-Beginners} (applying the fix from \cite{ChrombiumWebRTCBug}).

\iflong
\subsubsection{Browser Distribution in Android}
\fi
\label{app:android2}
We want to estimate the browser market share of ``supportive'' browsers (Chrome-like and Firefox) in the Android OS.
Based on April 2018 figures for operating systems,\footnote{\url{https://netmarketshare.com/operating-system-market-share.aspx}} combined with 
 mobile browsers distribution in April 2018,\footnote{\url{https://netmarketshare.com/browser-market-share.aspx}} 
 we conclude that the Chrome-like browsers (Google Chrome, Opera Mini, Baidu, Opera) comprise 90\% of the browser usage in Android. Adding Firefox (even though its STUN traffic is less bursty, Firefox can still be tracked at least for $f=100$) gets this figure up to 92\%. 

\iflong
\subsubsection{Chrome's STUN Traffic Shape}
\fi
\label{app:android3}
\label{subsec:chrome-stun}
Chrome sends the STUN requests to the list of supposedly STUN servers, in bursts. A single burst may contain the full list of the requested STUN servers (in ascending order of destination IP address), or a subset of the ordered list (typically with a missing range of destination hosts).
We measured 1014 bursts (to $L=400$ destination IP addresses) emitted by a Google Pixel 2 mobile phone (Android 8.1.0, kernel 4.4.88), running Google Chrome 67 browser. The vast majority of bursts last between 0.1 seconds to 0.2 seconds, and the maximal burst duration was 0.548 seconds. Thus we use an upper bound of $\delta_L=0.6$ seconds for a single burst duration.

Chrome emits up to 9 bursts with increasing time delays, at the following times (in seconds, where $t=0$ is the first burst): 0, 0.25, 0.75, 1.75, 3.75, 7.75, 15.75, 23.75, 31.75.\footnote{See \url{https://chromium.googlesource.com/external/webrtc/+/master/p2p/base/stunrequest.cc}).} We label these bursts $B_0,\ldots,B_8$ respectively, and we will be interested in $B_4$ and $B_5$, as they're sufficiently far from their neighbors. Thus, we are only interested in the first 8-9 seconds of the STUN traffic.

\iflong
\subsubsection{UDP Latency Distribution}
\fi
\label{app:android4}
While WebRTC traffic is emitted by the browser in well defined, ordered bursts, one cannot assume the traffic will retain this ``shape'' when arriving to the destination servers. Indeed, even order among packets within a burst is not guaranteed at the destination. Understanding  the latency distribution in UDP short datagrams is therefore needed in order to simulate the in-the-wild behavior, and consequently the efficacy of various tracking techniques.
The latency of UDP datagrams is gamma-distributed according to \cite{Li2007EstimationAS} and \cite{MaheshwariVMK17}. However, for simplicity, we use normal distribution to approximate the in-the-wild latency distribution. On May \nth{1}-\nth{6} 2018, we measured the latency 
of connections
to a  server in Microsoft Azure ``East-US'' location (in Virgina, USA) 
from 8 different networks located in Israel, almost 10,000km away.
The maximum standard deviation was 0.081 seconds. Hereinafter, we will use a standard deviation value $\sigma=0.1$ seconds as a worst case scenario for UDP jitter.

\iflong
\subsubsection{Packet Loss}
\fi
\label{app:android5}
We identified two different packet loss scenarios:
\begin{itemize}
\itemshort Packet loss during generation: the WebRTC packet stream (in Chrome-like browsers) is bursty in nature. In some bursts, we noticed large chunks of missing packets. These are quite rare (in the  
\iflong
experiment we describe in Section~\ref{subsec:chrome-stun}
\fi
we got 29 such cases out of 1014 -- 2.9\%, though they are more common in Androids whose kernel is 4.x and have $f=100$) and easily identified. We can safely ignore them because the tracker can detect a burst with a lot of missing packets,  reject the sample and run the  sampling logic again, or use a more sophisticated logic incorporating information from more than two bursts. Additionally, with $f=100$ there are far less false pairs, which helps the analysis.
\itemshort Network packet loss: the UDP protocol does not guarantee delivery, and indeed packets get lost over the Internet. The loss rate is not high, however, and we estimate it to be $\leq 1\%$. This is also backed by research.\footnote{See \url{http://www.verizonenterprise.com/about/network/latency/}, and  \cite{Baltrunas}.}

\end{itemize}

\subsection{The Tracking Technique}
\label{sec:linuxattack}
The technique that we use is different than prior art techniques in focusing on bucket {\em collisions}. That is, in cases wherein UDP datagrams for two different destination IP addresses end up with IPID generated using the same counter.

The tracker needs to control $L$ Internet IPv4 addresses, such that the IP-level traffic to these addresses (and particularly, the IP ID field) is available to the tracker. Ideally the IPs are all in the same network, so that they are all subject to the same jitter distribution. The tracker should be able to monitor the traffic to these IP addresses with time synchronization resolution of about 10 milliseconds (or less) - e.g. by having all the IPs bound to a single host.

With $L$ different destination IP addresses and $M$ buckets ($M=2048$ in Algorithm~\ref{alg-linux-ipid}), there are $\nicefrac{\binom{L}{2}}{M}$ expected
\iflong
collisions (unordered pairs of IP addresses which fall to the same bucket), 
\fi
assuming no packet loss.
In reality, the tracker can only obtain an {\em approximation} of this set. The goal is to reduce those false negatives and false positives to levels which allow assigning meaningful tracking IDs.

The basic property that enables the attacker to construct the approximate list is that in an IP ID generation the counter is updated by a random number which is smaller than 1 plus the multiplication of the timer frequency $f$ and the time that passed since the last usage of that counter.  Therefore for a true pair  $(\IP^i,\IP^j)$ where the IP ID generation for $\IP^i$ and $\IP^j$ used  the same bucket (counter), the following inequality almost always holds:
$$0<(\IPID^j-\IPID^i) \mod 2^{16}< f\Delta t+10$$
(We use $f\Delta t+10$ instead of $f\Delta t+1$ to support up to 10 IPs colliding into the same bucket, as each collision may increment the counter by $\leq 1+f\Delta t$ where $\Delta t$ is from the {\em previous} collision. So the counter can end up incrementing no more than $f \Delta t + 10$ where $\Delta t$ is the sum of the time difference between collisions, i.e. the time duration between the first collision and the last collision in the burst.)

Since we are looking at datagrams from the same burst we have an upper bound $\delta_L$  such that $\Delta t<\delta_L$, and therefore:
$$0<(\IPID^j-\IPID^i) \mod 2^{16}< f\delta_L+10$$

For two IP addresses which are \textit{not} mapped to the same counter,  the likelihood of this inequality to hold is only $\frac{(f\delta_L+10)-1}{2^{16}}$ which is 
\iflong
$\ll 1$ when $f\delta_L \ll 2^{16}$
(the worst case in our setting is with $f=300$ and $\delta_L=0.6$, where  $\frac{(f\delta_L+10)-1}{2^{16}}< 0.003$).
\fi
The key extraction algorithm (Section~\ref{sec:exhaustive-search}) will examine  IP ID values in two different communication bursts, and this will further reduce the likelihood of a false positive. Note that the probability of a false positive pair in a given burst to survive into the next burst is roughly $\frac{f\delta_L+10}{f\Delta t} \approx \frac{\delta_L}{\Delta t}$ where $\Delta t$ is the time between the consecutive bursts, whereas a true pair will occur in all bursts. Thus for the intersection of 2 consecutive bursts $\Delta t=4$ seconds apart, the amount of false positives (in both bursts) will be $\approx 0.15$ of their amount in a single burst.

\iflong
Another requirement is that the set $L$ of IP addresses is large enough, so that the number of colliding pairs will be sufficiently high in most of the tracking attempts, rather then on the average (the {\em expected} number of colliding pairs is $\nicefrac{\binom{L}{2}}{M}$). 
\fi

\subsection{Attack Phase 1 -- Collecting Collisions}
\label{sec:linux-collecting-collisions}
The tracking snippet needs to be rendered for at least 8.5 seconds, enough time for the browser to send the first 6 STUN bursts ($B_0,\ldots,B_5$) -- see Section~\ref{subsec:chrome-stun}. 
The tracking server splits the STUN traffic to bursts, based on the datagrams' time of arrival, and on the expected burst time offsets (see Section~\ref{subsec:chrome-stun}). For simplicity and ease of analysis, we henceforth only use traffic from bursts $B_4$ and $B_5$, which can be easily and unambiguously determined (since they are well separated in time from other bursts). We note that in some cases, requests in $B_4$ or in $B_5$ may be unsent, and in such cases we may need to resort to using e.g. $B_3$ and $B_5$ or similar combinations, but as long as these are ``late'' bursts (i.e. separated from their neighboring bursts by a  enough $\sigma$ units, where $\sigma$ is the UDP jitter, see above), they can be separated without errors (or almost without errors) and the following analysis remains valid.
If there are too many missing requests in a burst, the Tracking Server communicates with the Tracking Snippet, instructing it to retest the device. 

Assuming no (or few) missing requests in $B_4$ and $B_5$, the Tracking Server starts analyzing the data per burst (in $B_4$ and $B_5$). 
For each burst the Tracking Server calculates a set of pair candidates by collecting pairs of IP addresses
\iflong
\footnote{Chrome-like browsers send STUN requests ordered by IP address, thus for a true pair a higher IP will have a higher IP ID.}  
\fi
$(IP^i,IP^j)$ for which  $IP^i<IP^j$ and $0<(IPID^j-IPID^i) \mod 2^{16} < \lambda_L$ where $\lambda_L=f\delta_L+10$.
It then identifies pairs which appear in the candidate sets of \textit{both} bursts, and adds them to a set $U$  of full candidates. This set forms a single {\em measurement} of a device.
The tracker calculates the tracking ID based on $U$ in Phase 2. 
\subsection{Attack Phase 2 -- Exhaustive Key Search}
\label{sec:exhaustive-search}
In the second phase the tracking server runs an exhaustive search on the key space $W$ where the key is 32 bits long for algorithms $A_1$ and $A_2$, 41 bits long for algorithm $A_3$ (Linux) and 48 bits for $A_3$ (Android). For each candidate key, the algorithm counts how many IP pairs in $U$ are predicted by the candidate key. It is expected that only in one (the correct) key, this number will exceed a threshold $\nu$, and in such case, this will be returned as the correct key (and the device ID). See Algorithm \ref{alg-variant2} for details (the algorithm uses the notation $h'(...,k)=h(...,protocol\oplus g(net),key)$ where $k$ is split into $g(net),key$).

\begin{algorithm}
\caption{Exhaustive key search}
\label{alg-variant2}
\begin{algorithmic}[1]
\Procedure{Generate-ID}{$U, IP_{src}$}
\Comment{$U$ is defined in Section \ref{sec:linux-collecting-collisions}}
\If{$|U|<\nu$} 
\State return ERROR  
\EndIf
\State $X \gets \emptyset$
\ForAll{$0 \leq k < W$}
\State $Y \gets \{(IP^i,IP^j) \in U \;|\; h'(IP^i,IP_{SRC},k)  = h'(IP^j,IP_{SRC},k)\}$
\If {$|Y| \geq \nu$}
\State $X \gets X \cup \{k\}$
\EndIf
\EndFor
\If {$|X|>0$}
\State return $X$ \Comment{Needs special treatment if $|X|>1$}
\Else
\State return ERROR
\EndIf
\EndProcedure
\end{algorithmic}
\end{algorithm}
We assume here knowledge of the version of the algorithm ($A$) used -- $A_1$, $A_2$ or $A_3$. For $A_1$ and $A_2$, the key space size is $|W|=2^{32}$, and for $A_3$, it is $2^{41}$ for the x64 architecture and $2^{48}$ for the ARM64 architecture (see Section~\ref{sec:keyspace}.)

As explained in Section \ref{sec:linux-practical-considerations}, false positives ($|X|>1$) are very rare -- they can be handled but as this complicates the analysis logic, it is left out of the paper.

\mypar{Attack run time} Where $|U|=P$ pairs, the run time of Algorithm~\ref{alg-variant2} is proportional to $|W|P$. $P$'s distribution depends on $f$; Table~\ref{Table:P-distribution} summarizes the expectancy and standard deviation for common $f$ values. These were approximated by a computer simulation (100 million iterations.)
\begin{table}[]
\caption{Approximated $P$ distribution}
\label{Table:P-distribution}
\begin{center}
\begin{tabular}{|l|l|l|l|}
\hline
$f$ [Hz] & $E(P)$ & $\sigma(P)$ & $\nicefrac{\sigma(P)}{E(P)}$ \\ \hline
100 & 50.59 & 7.39 & 0.146\\ \hline
250 & 65.47 & 8.60 & 0.131\\ \hline
300 & 70.45 & 8.79 & 0.125\\ \hline
\end{tabular}
\end{center}
\end{table}
\mypar{Time/memory optimization}
When the number  of devices to measure is much smaller than $|W|$ it is possible to optimize the technique for repeat visits. The optimization simply amounts to keeping a list $\Lambda$ of already encountered $key$  values (or $(g(net),key)$  values), and trying them first. If a match is found (i.e., this is a repeat visit), there is clearly no need to continue searching the rest of the key space. 
Otherwise, the algorithm  needs to go through the remaining key space. 
\mypar{Targeted tracking}
Even if the key space $W$ is too large to make it economically efficient to run large scale device tracking, it is still possible to use it for targeted tracking. The use case is the following:
The tracking snippet is invoked for a specific target (device), e.g. when a suspect browses to a honeypot website. At this point, the tracker (e.g. law enforcement body) extracts the key, possibly using a very expensive array of processors, and not necessarily in real time. Once the tracker has the target's key, it is easy to test any invocation of the tracking snippet against this particular key and determine whether the connecting device is the targeted device. Moreover, if the attacker targets a single device (or very few devices), it is possible to reduce the number of IP addresses used for re-identifying the device, by using only IP addresses which are part of pairs that collide (into the same counter bucket) under the known device key. Thus we can use a single burst with as few as 5 IP pairs 
\iflong
(10 addresses altogether)
\fi
per device to re-identify the device. The dwell time in this case drops to near-zero.

\subsection{{The Effective Key Space in Attacking Algorithm \it A\textsubscript{3}}}
\label{sec:keyspace}
In Algorithm $A_3$, 32 bits of the net namespace are extracted by a function we denote as  $g()$, and are added to the calculation of the hash value. 
The attack depends on the effective keyspace size $|W|=|\{key\}| \times |\{g(net)\}|=2^{32}\cdot |\{g(net)\}|$.  

\iflong
A detailed analysis for Linux kernel versions 4.8 and above (on x64), and 4.6 and above (on ARM64, i.e. Android) appears in Appendix~\ref{app:netspace}. The conclusion is 
\fi
that if KASLR is turned off then the effective key space size is 32 bits in both x64 and ARM64. If KASLR is turned on, then the effective key space size is 41 bits in x64 and 48 bits in ARM64.

\subsection{KASLR Bypass for Algorithm \it A\textsubscript{3}}
By obtaining $g(net)$ as part of Attack Phase 2 (Section \ref{sec:exhaustive-search}), the attacker gains 32 bits of the address of the {\tt net} structure. In single-container systems
such as desktops and mobile devices, this {\tt net} structure resides in the {\tt .data} segment of the kernel image, and thus has a fixed offset from the kernel image load address. 
In default x64 and ARM64  configurations, the 32 bits of $g(net)$ completely reveal the random KASLR displacement of {\tt net}. This suffices to reconstruct the kernel image load address and thus fully bypass KASLR. 
\iflong
See Appendix~\ref{app:netspace} for more details.
\fi

\subsection{Optimal Selection of $L$}
\label{App:Linux-optimal-L}
Since IP addresses  are at premium, we choose a minimal integer  number $L$ of IP addresses such that at the point $\nu$ where $Prob(FN)+Prob(FP)$ is minimal, $Prob(FP)+Prob(FN) \leq 10^{-6}$. We assume $f=300$ (worst case scenario). For simplicity, at this stage we neglect packet loss, and  assume that $\delta_L=\frac{L}{400}\delta_{400}$ (we assume $\delta_L \propto L$, and we measured $\delta_{400}$). For false negatives, we use the Poisson approximation of birthday collisions \cite{arratia1989} with $\lambda={\binom{L}{2}}/M$. Therefore:
$$Prob(FN) \approx \sum_{i=0}^{\nu-1}\frac{\lambda^i e^{-\lambda}}{i!}$$
For false positives, we also assume that a burst contains the average number of false pairs and true pairs  $A=\lfloor \frac{f\delta_L+10}{f\Delta t}\binom{L}{2}\frac{f\delta_L+10}{2^{16}}+\frac{\binom{L}{2}}{M} \rfloor$. 
\iflong
We note that the probability for a single false key to match exactly $k$ pairs is $\binom{A}{k}(\frac{1}{M})^{k}(1-\frac{1}{M})^{A-k}$, 
thus the probability of a single false key not to become a false positive is $\sum_{i=0}^{\nu-1}\binom{A}{i}(\frac{1}{M})^{i}(1-\frac{1}{M})^{A-i}$.
\fi
The probability of $|W|-1$ false keys to generate at least one false positive key is therefore:
$$Prob(FP) \approx 1-\Big(\sum_{i=0}^{\nu-1}\binom{A}{i}(\frac{1}{M})^{i}(1-\frac{1}{M})^{A-i}\Big)^{|W|-1}$$
Assuming $|W|=2^{48}$ (worst case -- Android), we enumerated over all $\nu$ values for each $L$ in $\{200,250,\ldots,500\}$ to find the optimal $\nu$ (per $L$). 
\iflong
We plot the results in Fig. \ref{L-graph}. 

\begin{figure}[t]
\caption{$Prob(FP)+Prob(FN)$ for $L$ and optimal $\nu$}
\includegraphics[width=8cm]{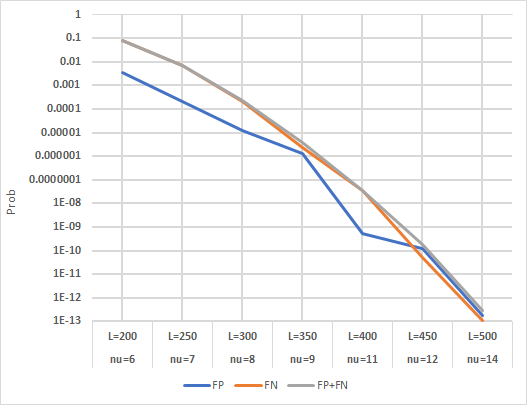}
\label{L-graph} 
\end{figure}

\fi
\iflong
As can be seen, 
\fi
$L=400$ (with $\nu=11$) is the minimal ``round'' $L$ satisfying $Prob(FP)+Prob(FN) \leq 10^{-6}$ at its optimal $\nu$.

\subsection{A More Accurate Treatment for $L=400$}
\label{App:FP-FN-L400}
Using a computer simulation, we approximated the distributions of all collisions  $p_A(n)$ 
(using $10^{8}$ simulation runs),
and of true collisions $p_T(n)$ (using $10^9$ simulation runs). The simulations took into account 1\% packet loss. With these, we can calculate more accurate approximations:
$$Prob(FN) \approx \sum_{i=0}^{\nu-1}p_T(i)$$
$$Prob(FP) \approx 1-\sum_{n} p_A(n)\Big(\sum_{i=0}^{\nu-1}\binom{n}{i}(\frac{1}{M})^{i}(1-\frac{1}{M})^{n-i}\Big)^{|W|-1}$$
(We use the convention $\binom{n}{k}=0$ where $k>n$.) We enumerated over values $1 \leq \nu \leq 20$ for $L=400$ and $|W|=2^{48}$ (worst case -- Android.) 
\iflong
and plotted the FP and FN probabilities in Fig.~\ref{nu-graph} (eliminating values below $10^{-15}$). 

\begin{figure}[t]
\caption{$Prob(FP)+Prob(FN)$ vs. $\nu$ for $L=400$}
\includegraphics[width=8cm]{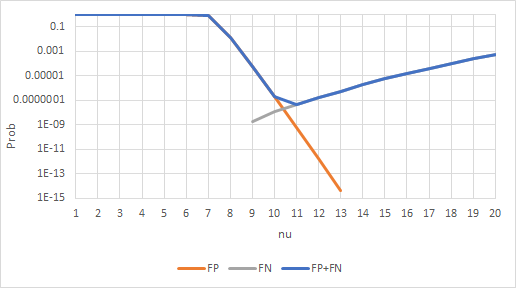}
\label{nu-graph} 
\end{figure}

As can be seen, the
\fi
minimal $Prob(FP)+Prob(FN)$ is at $\nu=11$, where $Prob(FP)=6.2 \times 10^{-10}$ and $Prob(FN)=4.2 \times 10^{-8}$. We get the same optimal $\nu$ value for $L=400$ as we got in Section~\ref{App:Linux-optimal-L}, which means that the approximation steps we took there are reasonable.

\subsection{Practical Considerations}
\label{sec:linux-practical-considerations}
\paragraph{Controlling packets from the browser}
As explained in Section~\ref{sec:STUN}, it is possible to emit UDP traffic to arbitrary hosts and ports using WebRTC. The packet payload is not controlled. The tracker can use the UDP destination port in order to associate STUN traffic to the same measurement. 

\mypar{Synchronization and packet transmission/arrival order}
Unlike the Windows technique, in the Linux/Android tracking technique there is no need to know the exact transmission order of the packets within a single burst. 
\mypar{False positives and false negatives}

Using a computer simulation with $L=400$ destination IP addresses, a burst length of $\delta_L=0.6$ seconds, and packet loss rate of 0.01, we calculated an approximation  of
for the false negative rate of $4.2 \times 10^{-8}$ for $\nu=11$, and an approximation for the false positive rate of $6.2 \times 10^{-10}$. These approximations were computed  assuming $|W|=2^{48}$ (worst case -- Android). See Section~\ref{App:FP-FN-L400} for more details. 
\mypar{Device ID collisions}
The expected number of pairs of devices with colliding  IDs, due to the birthday paradox,  and given $R$ devices and a key space of size $|W|$, is $\nicefrac{\binom{R}{2}}{|W|}$. 
For Algorithms $A_1$ and $A_2$ the key space size is $|W|=2^{32}$, and will cause device ID collisions  once there are several tens of thousands of devices. For $R=10^6$ this will affect 0.00023 of the population (2 out of every 10,000 devices). 
For Alg. $A_3$, the key space size (with KASLR) is $\geq 2^{41}$, so collisions  start showing up with $R$ in the millions. Even for  $R=128\cdot 10^6$,   collisions  affect only 0.00006 of the population.
\mypar{Dwell time}
In order to record $B_5$, the snippet page needs to be loaded in the browser for 8-9 seconds. Navigating away from the page will immediately terminate the STUN traffic. 
\mypar{Environment factors}
All the UDP-related topics in Section \ref{sec:env-factors} are applicable as environment factors on the Linux/Android tracking technique. 
\mypar{Longevity}
The device ID remains valid as long as the device is not shutdown or restarted. Mobile devices are rarely shut down, and are typically restarted only on updates, which happen once every several months, or even less frequently.
\mypar{Scalability}
The attack is scalable. Device ID collisions are rare even with many millions of devices (see above). False positives and false negatives are also rare (less than $4.3 \times 10^{-8}$ combined). From a resource perspective, the attack uses a fixed number of IPs and servers, and a fixed-size RAM/disk. The required CPU power is proportional to the number of devices measured per time unit. Network consumption per test is negligible -- approx. 13.5KB/s (at the IP level) during measurement.
\subsection{Possible Countermeasures}
\label{Sec:linux-mitigation}
\paragraph{Increasing {\it M}}
Changing the algorithm to use a larger number  $M$ of counters, will reduce the likelihood of pairs of IP addresses using the same counter. 
In  response to such a change the tracker can increase the number  $L$ of IP addresses that is uses. 
The  expected number of collisions is $\nicefrac{\binom{L}{2}}{M}$, and therefore increasing  $M$ by a factor of $c$  requires the attacker to increase $L$ by only a factor of $\sqrt{c}$.

On the other hand,  $\delta_L$ also grows (probably linearly in $L$), and when $f\delta_L \geq 2^{16}$ no information is practically revealed to the tracker. It is probably safe to assume that the tracker can handle an increase of  $L$  by a factor of $\times 10$, which means that in order to stop the attacker the IP ID generation algorithm must increase  $M$ by  more than $\times 100$, making it too memory expensive to be practical.
\iflong
(Decreasing $M$ only makes it easier for the tracker, except for $M=1$ in which case the system becomes a global counter (with random hops depending on the time since last transmission), which renders the attack ineffective, yet has several other security issues, such as described for example in~\cite{idlescan}.)
\fi
\mypar{Increasing the key size ({\it W})}
This can be an effective counter-measure for the exhaustive search phase, though the pair collection phase is unaffected by it. Yet some choices of the hash function  $h$ might still allow fast cryptanalysis.
\mypar{Strengthening {\it h}}
Our analysis does not rely on any property of the hash function $h$, except that it is more-or-less uniform. Thus, changing $h$
\iflong
(while keeping the same key space $W$) 
\fi
will not affect our results.
\mypar{Replacing the algorithm}
See the last item in Section~\ref{Windows-mitigation}.

\section{Experiment -- Attacking Linux and Android Devices in the Lab}
\label{Sec:Linux-Experiment}
In order to verify that we can extract the key used by  Linux and Android devices, we need to control hundreds of IP addresses. 
Controlling such a magnitude of Internet-routable IP addresses was logistically out of scope for this research. Therefore we had to settle for an in-the-lab setup, which naturally limited the number of devices we could test.
\subsection{Setup}
We connected the tested devices to our own WiFi access point, which advertised our laptop as a network gateway. Then we launched a Chrome-like browser inside the Linux/Android device, and navigated to a page  containing a tracking snippet. The tracking snippet used WebRTC to force UDP traffic to a list of $L=400$ hosts, and this traffic passed through our laptop (as a gateway) and was recorded.

We then ran the collision collection logic (Phase 1), and fed its output (IP pairs whose IP IDs collide) to the exhaustive key search logic (Phase 2). For KASLR-enabled devices, we also provided the algorithm with the offset (relative to the kernel image) of {\tt init\_net}, which we extracted from the kernel image file given the build ID (can be inferred e.g. from the {\tt User-Agent} HTTP request header).
We expected that the algorithm will output a single key, which will match a large part of the collisions.

\subsection{Results}
\label{Sec:Linux-results}
We tested 2 Linux laptops and 6 Android devices, together covering the vast majority of operating system and hardware parameters that regulate the IP ID generation. The results from all tests were positive - our technique extracted a single key and a kernel address of {\tt init\_net} where applicable (which was identical to the address in {\tt /proc/kallsyms}). Note that due to hardware availability constraints, for the Pixel 2XL case ($|W|=2^{48}$), we provided the algorithm with the correct 16 bit kernel displacement to reduce the key search to $2^{32}$. Table~\ref{table:linux-combinations} provides information about the common kernel versions, their parameter combinations and the tested devices. 

The Attack time column is the extrapolated attack time in seconds with 10,000 Azure B1s machines, based on $E(P_f)$ from Table~\ref{Table:P-distribution}, i.e. the average attack time is $r \cdot |W| \cdot E(P_f)$ where $r$ is the time it takes a single B1s machine to test a single key with a single pair, divided by 10,000. The standard deviation of the attack time for a given $f$ is $r \cdot |W| \cdot \sigma(P_f)$, which is $\nicefrac{\sigma(P_f)}{E(P_f)}$ in Table~\ref{Table:P-distribution} times the average attack run time in Table~\ref{table:linux-combinations}.
From a calibration run (single B1s machine, 10 pairs, $2^{32}$ keys, 294.83 seconds run time) we calculated $r=6.8645\times 10^{-13}$, and populated the Attack Time column in Table~\ref{Table:P-distribution} with $r \cdot |W| \cdot E(P_f)$.

\begin{table*}[t]
\caption{Common Linux/Android Kernels and Their Parameter Combinations}
\label{table:linux-combinations}
\begin{tabular}{|l|l|l|l|l|l|l|l|l|l|}
\hline
O/S & \begin{tabular}[c]{@{}l@{}}Kernel\\ Version\end{tabular} & Alg. & $f$ [Hz] & KASLR & NET\_NS & $\rho$ & $\log_2|W|$ & Tested System & \begin{tabular}[c]{@{}l@{}}Attack\\ Time [s]\end{tabular} \\ \hline
Linux (x64) & 4.19+ & $A_3$ & 250 & Yes & Yes & 12 & 41 & \begin{tabular}[c]{@{}l@{}}Dell Latitude\\ E7450 laptop\end{tabular} & 99 \\ \hline
Linux (x64) & 4.8-4.18.x & $A_3$ & 250 & Yes & Yes & 6 & 41 & \begin{tabular}[c]{@{}l@{}}Dell Latitude\\ E7450 laptop\end{tabular} & 99 \\ \hline
\begin{tabular}[c]{@{}l@{}}Android\\ (ARM64)\end{tabular} & \begin{tabular}[c]{@{}l@{}}4.4.56+, \\ 4.9, 4.14\end{tabular} & $A_3$ & \begin{tabular}[c]{@{}l@{}}300/\\ 100\end{tabular} & Yes & Yes & 6/7 & 48 & Pixel 2XL ($\rho=6$)& \begin{tabular}[c]{@{}l@{}}13,612/\\ 9,775\end{tabular} \\ \hline
\begin{tabular}[c]{@{}l@{}}Android\\ (ARM64)\end{tabular} & \begin{tabular}[c]{@{}l@{}}3.18.17+\\ 3.4.109+\end{tabular} & $A_2$ & 100 & No & No & \begin{tabular}[c]{@{}l@{}}Don't\\ care\end{tabular} & 32 & \begin{tabular}[c]{@{}l@{}}Redmi Note 4\\ Xiaomi Mi4\end{tabular} & 0.15 \\ \hline
\begin{tabular}[c]{@{}l@{}}Android\\ (ARM64)\end{tabular} & \begin{tabular}[c]{@{}l@{}}3.18.0-3.18.6\\ 3.10.53+\\ 3.4.103-3.4.108\end{tabular} & $A_1$ & 100 & No & No & \begin{tabular}[c]{@{}l@{}}Don't\\ care\end{tabular} & 32 & \begin{tabular}[c]{@{}l@{}}Samsung J7 prime\\ Samsung S7\\ Meizu M2 Note\end{tabular} & 0.15 \\ \hline
\end{tabular}
\end{table*}    

\paragraph{Applicability in-the-wild}
While our tests were carried out in the lab, we argue that the results are representative of an in-the-wild experiment with the same devices. We list the following potential differences between in-the-lab and in-the-wild experiment, and for each difference, we note why our experiment can be projected to an in-the-wild scenario.
\begin{itemize}
\itemshort Packet loss: our technique is not sensitive to packet loss. We ran false positive/negative computer simulations (assuming 1\% packet loss) supporting this fact.
\itemshort Network latency: our technique is not sensitive to network latency (which is just a constant time-shift, from our perspective).
\itemshort UDP jitter: this only affects correctly splitting the traffic into bursts. Our technique uses the ``late'' bursts, thus assuring that the bursts are well separated time-wise and that a jitter of $\sigma=0.1$s does not affect tracking.
\itemshort Network interference (IPID modification): this issue was already evaluated in-the-wild in the Windows experiment, and the Windows results can be applied to the Linux/Android use case.
\itemshort Packet reordering (within a burst): Our technique does not rely on packet order within a burst.
\end{itemize}
Thus we conclude that our results (and henceforth, the practicality of our technique) are applicable in-the-wild.

\section{Conclusions}
\label{Sec:conslusions}
    Our work demonstrates that using  non-cryptographic random number generation of  attacker-observable values  (even if the values  themselves are not security sensitive), may be a security vulnerability in itself, due to an attacker's ability to extract the  key/seed used by the algorithm, and use it as a fingerprint of the system.
    
\iflong
Specifically, we find that the IP ID generation scheme proposed in RFC 7739  \cite[Section 5.3]{rfc7739}, and implemented in Windows 8 and later is vulnerable to an  attack of this type, and that the Linux/Android IP ID generation algorithm is also vulnerable to such an attack. We demonstrate these  attacks and are able to extract  device IDs for systems running Windows, Linux and Android.
\fi    
  
     We stress that any replacement cryptographic algorithm must not be hampered by using a key that is too short, in order to avoid a key enumeration attack.
     Also, as a security measure, we strongly recommend generating unique keys for such cryptographic usage, without resorting to using secret data that is used for other purposes (which -- in case of a cryptographic weakness in the algorithm -- can leak out).

\section{Acknowledgements}
This work was supported by the BIU Center for Research in Applied Cryptography and Cyber Security in conjunction with the Israel National Cyber Directorate in the Prime Minister's Office.

We would like to thank the anonymous
\iflong
USENIX 2019
\fi
 reviewers for their feedback, 
Assi Barak for his help to the project, as well as Avi Rosen, Sharon Oz, Oshri Asher and the Kaymera Team for their help with obtaining a rooted Android device.

\bibliographystyle{abbrv}


\appendix

\section{Details of the Attack on Windows}
\iflong
\subsection{Speeding up Phase 1 of the Windows Attack Using Preprocessing}
\label{app:windowspreprocessing}
Denote by $C \in GF(2)^{15(J-1) \times 30}$ the matrix of equation coefficients derived from $\IP^j \oplus \IP^0$, which is known at setup time. Entry 
$C_{15(j-1)+(i-17),(i-17)+(m-16)}$ in this matrix is equal to 
$(\IP^j \oplus \IP^0)_{m} $ for   $j\in [1,J-1],$ $i\in [17,31]$ and  $m\in [16,31]$, and to 0 elsewhere. 

\remove{
Denote by $C \in GF(2)^{15(J-1) \times 30}$ the matrix of equation coefficients derived from $\IP^j \oplus \IP^0$, which is known at setup time. The entries in this matrix are
$$
C_{15(j-1)+(i-17),(i-17)+(m-16)}=
\left\{ \begin{array}{ll}
 (\IP^j \oplus \IP^0)_{m} &  j\in [1,J-1], i\in [17,31], m\in [16,31] \\ 
 & \\
 0 & \mathrm{elsewhere} \\
 \end{array}
\right.
$$
}

During setup time, an ``inverse'' matrix, $Z \in GF(2)^{15(J-1) \times 15(J-1)}$ for $C$ is calculated using a Gaussian elimination process starting with
$\begin{pmatrix}
Id_{30} & 0 \\
0 & 0
\end{pmatrix}
\in GF(2)^{15(J-1) \times 15(J-1)}$
and $C$, so that \[
Z \cdot C = 
\begin{pmatrix}
Id_{30} \\
0
\end{pmatrix}
 \in GF(2)^{15(J-1) \times 30}
\]
For $j=1, \ldots, J-1$, define $D^j=(\Vectorize(\IPID^j-(\beta_0+j) \mod 2^{15}) \oplus \Vectorize(\IPID^0-(\beta_0) \mod 2^{15}))_{17,\ldots,31}$, then eq.~(\ref{eq:phase1-main}) becomes (for $i=17,\ldots,31$):
\begin{equation} \label{eq:pahse1-main-D}
    \bigoplus_{m=16}^{31}(\IP^j \oplus \IP^0)_{m} \cdot K_{i+m}=D^j_i
\end{equation}

Note that the most significant bit of $\beta_0 \mod 2^{15}$ cancels itself and does not affect $D^j$, hence it is not needed in order to calculate $D^j$ (i.e. the enumeration can take place over $\beta_0 \mod 2^{14}$). 
With $Z$ it is now easy to find the 30 bits of $K$: the tracker simply calculates $Z \cdot (D^0 || \cdots || D^J-1) $ to get a vector with the 30 $K$ bits, and $15(J-1)-30$ trailing zeros (if the guess of $\beta_0$ is correct). 
Note that if $15(J-1)$ is larger than the word size of the attacker CPU (e.g. 64), it is faster to use only as many equations as the word size for the matrix and vector operations, and then (with the known 30 bits of $K$ and eq.~(\ref{eq:pahse1-main-D})), proceed to use the remaining equations to eliminate keys.  
\subsection{Details of Phase 2 of the Attack}
\label{app:windowsphase2}
Note that the leftmost bit of $\beta_0 \mod 2^{15}$ cancels itself in  eq.~(\ref{eq:phase2-main}) (since $(x \oplus 2^{k-1}) \mod 2^{k} = (x+2^{k-1}) \mod 2^{k}$), therefore it is not needed in the calculation. Likewise,  $K_{17}$ cancels itself in eq.~(\ref{eq:phase2-main}) (it only appears as an {\em addendum} $(\IP^0_0 \oplus \IP^{g,j}_0)K_{17}$ in $T(K,\IP^0 \oplus \IP^{g,j})_{17}$, when expanded according to eq.~ (\ref{eq:T-def})), and its coefficient $\IP^0_0 \oplus \IP^{g,j}_0$ does not change between $j=0$ and $j=1$ because $\IP^{g,0}_0=\IP^{g,1}_0$ as they are in the same class B network). Therefore this bit is not needed as well.

\textit{Finding the remaining key bits:}
Thus, knowing $\beta_0 \mod 2^{14}$ and $K_{18},\ldots,K_{62}$ suffices to calculate the right hand side of eq.~(\ref{eq:phase2-main}). 
The values of $\beta_0 \mod 2^{14}$ and $K_{33},\ldots,K_{62}$
are known from phase 1, and since the left hand side is also known, the equation can be used to eliminate false guesses of $K_{18},\ldots,K_{32}$ (and also other false keys, such as ones caused by incorrect $\beta_0 \mod 2^{14}$). 

The bit $K_{18}$ deserves special attention. It appears only in $T(K,\IP^0 \oplus \IP^{g,j})_{17,\ldots,31}$ as $(\IP^0_1 \oplus \IP^{g,j}_1)K_{18}$ in the leftmost bit, and $(\IP^0_0 \oplus \IP^{g,j}_0)K_{18}$ in the second-from-left bit. If in all pairs it holds that  $\IP^0_0 = \IP^{g,j}_0$, then $K_{18}$ only appears (as an addendum) in the leftmost bit, which cancels itself in the subtraction. Thus, in order to extract $K_{18}$, the IP set needs to satisfy $\exists_{0<g \leq G} \IP^{g,0}_0 \neq \IP^0_0$. Define $Q=|\{g|\IP^{g,0}_0 = \IP^0_0\}|$, then this requirement is equivalent to $Q<G$. An optimal value of $Q$ is calculated in Appendix \ref{appendix:false-positives}.

\fi

\subsection{Practical Considerations}
\label{app:practical}
\subsubsection{Controlling Packets from the Browser}
\textbf{UDP:} As explained in Section~\ref{sec:STUN}, it is possible to emit UDP traffic to arbitrary hosts and ports using WebRTC. The packet payload is not controlled. The tracker can use the UDP destination port in order to associate STUN traffic to the same measurement. \\
\textbf{TCP:} WebSocket \cite{rfc6455} emits TCP traffic in a controlled fashion once a circuit is established, thus can be used by the snippet to fully control packet transmission. The downside of using TCP-based protocols is the TCP-level retransmission, which can introduce loss of synchronization between the device and the server side, regarding how many packets were sent.
The tracker can use the packet payload to mark packets that belong to the same measurement.

\subsubsection{Packet Transmission Order}
\label{Appendix:Windows-packet-order}
We encountered cases in the wild where the packet payload generation order is not identical to packet transmission order. Specifically, Microsoft IE and Edge are prone to this behavior. This is only relevant in the same class B network (since there the original extraction algorithm makes an assumption on the order of the packets). Therefore, the tracker should try all possible permutations of packet order (per class B network IPs). In Phase 1, this means enumerating over all $\pi \in S_J$ ($J!$ permutations). For each permutation, use  the following definition of $D^j$ (instead of the original one):
$$D^j=(\Vectorize(\IPID^j-(\beta_0+\pi(j)) \mod 2^{15}) \oplus$$ $$\Vectorize(\IPID^0-(\beta_0+\pi(0)) \mod 2^{15}))_{17,\ldots,31}$$
\iflong
Observe that the IPID values used in $D^j$ are in the same order of the IP addresses used to calculate $Z$, therefore there is no need to re-calculate $Z$ with each iteration. 
\fi
It follows that enumerating over all possible orders will increase the run time of Phase 1 by a factor of $J!$. 
In Phase 2, for each pair of IP addresses, there are only $2!$ permutations, and since the elimination is so powerful, this will only affect the run time due to the first pair, i.e. will double the run time. 
\subsubsection{Handling False Positives}
\label{appendix:handling-false-positives}
The issue of false positive keys is covered in 
\iflong
Appendix \ref{appendix:false-positives}. 
\fi
As mentioned there, the vast majority of false positive keys only differ from the correct key by a few leftmost bits.
Table \ref{Table:common-tail} 
demonstrates that with an optimal choice of 30 IP addresses, if the tracker keeps only 41 bits of the key tail, he/she will get multiple keys with probability $2.1 \times 10^{-6}$, which is  sufficiently small even  for a large scale deployment.

\iflong
\begin{table}
\caption{Common tail length probability - measured with 1000 randomly chosen sets of 30 IPs ($J=6, G=12, Q=3$), 10,000 tests each ($10^7$ tests altogether).}
\label{Table:common-tail}
\begin{center}
\begin{tabular}{|l|l|}
\hline
\begin{tabular}[c]{@{}l@{}}Common\\ tail [bits]\end{tabular} & Prob.    \\ \hline
45                                                      & 0.9937579 \\ \hline
44                                                      & 0.0058328 \\ \hline
43                                                      & 0.0003783 \\ \hline
42                                                      & $2.6 \times 10^{-5}$  \\ \hline
41                                                      & $2.9 \times 10^{-6}$ \\ \hline
$\leq 40$                                                    & $2.1 \times 10^{-6}$ \\ \hline
\end{tabular}
\end{center}
\vspace{-0.5cm}
\end{table}

\fi
In the case where multiple keys are emitted by the algorithm (even after truncation, e.g. to 41 bits), two strategies can be applied: either (a) determining that this particular device cannot be assigned an ID (at the price of losing $2.1 \times 10^{-6}$ of the devices); or (b) assigning multiple IDs to the device (which makes tracking the device more complicated and more prone to ID collisions). 
\subsubsection{Handling False Negatives and Interference}
\label{App:Windows-false-negatives}
It is important to note that while there may be false positives, there are no algorithmic false negatives, i.e. the algorithm always emits the correct key (possibly along with incorrect keys), given the correct data. However, it is possible for the algorithm to receive incorrect data, in the sense that IP IDs are provided which are not (even after re-ordering) derived from an incrementing counter -- i.e. there are ``gaps'' in the counter values associated with the IP IDs. This can happen either due to packet loss or due to interference.\\
\textbf{Packet loss:} In TCP, when a packet from the client to the server is lost, the client will note a missing ACK and will eventually retransmit the original data (with incremented IP ID). This will cause a gap in the counter values, which can be enumerated by the analysis logic (our analysis logic does not currently implement this). Another packet loss scenario is wherein the ACK packet from the server to the client is lost, and  the client retransmits the original data. The server however receives two such packets, and can simply discard the one with incremented IP ID. Thus the ``problematic'' scenario is the one wherein the original data packet is lost.\\
\textbf{Interference:} Theoretically, an unrelated packet sent by the Windows device in between measurement packets may interfere with the measurement. However, this is very unlikely. First, the interfering packet must fall into the same counter bucket as that of the measurement packets -- this happens with probability of $\nicefrac{1}{8192}$ for a given bucket. Second, the timing is delicate -- the interfering packet should be sent between the first and the last measurement packet. This time window is below 1 second, so overall the likelihood for interference is very low. When such an interference happens, it creates a gap in the IP ID values, which can be addressed as explained below.\\
\textbf{Addressing ``gaps'':} The analysis logic can compensate for up to $l$ lost packets in the first class B network ($g=0$) by enumerating over all possible $\sum_{d=0}^l \binom{J-2+d}{J-2}$ gap configurations (each addendum counts all weak compositions of $d$ into $J-1$ parts). In our experiment, we measured $l=4$ for some ``difficult'' networks, so for $J=6$ there are 126 gap configurations, thus a $\times 126$ factor in the runtime. Note that a gap in the transmissions for $g>0$  is much easier to handle, as $(\IPID^{g,1}-\IPID^{g,0})$ in eq~(\ref{eq:phase2-main}) may now take values in $\{1,\ldots,l+1\}$, so there is no runtime factor for this case.
When the total gap space is larger than $l$, the algorithm will yield no key. In such a case, the server can instruct the snippet to run another test. 
Therefore the actual false negative probability can be reduced to as small as necessary.

\iflong
\subsection{Deploying the Attack when Javascript is Disabled}
\label{app:nojavascript}
When Javascript (or any client side scripting) is not available, e.g. with NoScript browser extension \cite{noscript}, it is still possible to send TCP packets to arbitrary hosts. Instead of orchestrating the packet transmission with Javascript, the snippet simply instructs the browser to load images (via {\tt IMG} HTML tag) from the destination IP addresses. This generates TCP traffic to the IP addresses, so it should be still possible for the tracking host to reconstruct a device ID from the key $K$. It is of course a bit more complicated since (unlike the Javascript-based implementation), the increments in $\beta[i]$ between packets sent to different IPs are no longer 1 - they rely on the exact order of transmission, and also need to take into account SYN packets, ACK packets, and FIN packets. 
\fi

\iflong
\subsection{Speeding the Attack Using Incremental Enumeration}
\label{sec:incrementalenumeration}
In the first phase of the attack, instead of enumerating over $2^{14}$ values of $\beta_0$ en-masse, it is possible to enumerate over the least significant 3 bits, solve the linear equations to yield 16+3  bits of $K$ (with $4(J-1)$ elimination bits), and for the surviving guesses, proceed to guess the next $K$ bit and the next $\beta_0$ bit,  eliminate using additional $J-1$ bits, and so on. The second phase of the attack proceeds along similar lines, with the remaining bits of $K$ guessed one by one and eliminated.
We did not implement this improvement as the straightforward implementation described above sufficed to ensure real-time tracking.
\fi

\iflong
\section{Assuring that the Matrix $C$ in the Windows Attack is Invertible}
\label{appendix:rank-ker-C}
We discuss here necessary conditions for the condition that $\rank(\ker(C))=0$ to hold. (This condition is required in order for the matrix $C$ to be invertible and allow  learning sufficient linear equations for extracting the key in the first phase of the attack.) 

If $C$ was chosen at random from $GF(2)^{15(J-1) \times 30}$, then per \cite{zbMATH02681320}, $\Pr(\rank(C)=30)=\prod_{a=0}^{29}(1-\frac{1}{2^{15(J-1)-a}})$, and for $J=6$ this yields $\Pr(\rank(\ker(C))>0) \approx 2^{-45}$.

However, in our case $C$ has a special structure, very unlike a random matrix, which affects the distribution of $\rank(\ker(C))$. It is therefore needed to check the probability of this event based on this distribution. 

We ran an experiment where we simulated 10,000,000 sets of destination IP addresses. The following are the lead factors with regards to the invertability of $C$, according to our experiments: 
\begin{itemize}
\itemshort Denote by $n_1$ the fixed {\em first} common bits inside the class B subnet of all $\IP^j$ (i.e. the first $16+n_1$ bits of the IP address are fixed). Since (trivially) $\rank(\ker(C)) \geq n_1$, it follows that $n_1=0$ is a necessary condition for $\rank(\ker(C))=0$. 
\itemshort Likewise, denote by $n_2$ the fixed {\em last} common bits inside the class B subnet of all $\IP^j$. Since $\rank(\ker(C)) \geq n_2$, it follows that $n_2=0$ is a necessary condition for $\rank(\ker(C))=0$. 
\itemshort Due to the specific structure of the equations, there are some (cyclic) nonzero vectors in $\ker(C)$ given some simple (but rare) conditions on $\IP^j$. The most common condition is $\forall_{0<j\leq J-1}\bigoplus_{m=16}^{31}(\IP^j \oplus \IP^0)_{m}=0$ which, if met (probability for random $\IP^j$ values is $2^{-(J-1)}$), yields $C \cdot (1,1,\ldots,1)=0$ (i.e. this vector of cycle length 1 is in $\ker(C)$). Thus, a necessary condition for $\rank(\ker(C))=0$ is $\exists_{0<j\leq J-1}\bigoplus_{m=16}^{31}(\IP^j \oplus \IP^0)_{m}=1$. Other such vectors exist for longer cycles, but their conditions are much more rare. 
\end{itemize}
IP sets of size $J=6$ that fulfill the above three requirements have a probability of roughly 1:1000  to yield $\rank(\ker(C))>0$ (We estimated this probability empirically, by simulating 10,000,000 random sets, of which  10,257 yielded $\rank(\ker(C))>0$). 
This demonstrates that by fulfilling these three simple requirements (which are indeed very easy to meet), $C$ becomes ``invertible'' with very high probability (99.9\% for $J=6$). Of course, to determine whether a particular $C$ is ``invertible'' requires applying the Gaussian elimination process (and calculating $Z$).

\textbf{Conclusion:} There should not be any problem choosing an IP set that makes $C$ ``invertible'' (in setup time, before starting to deploy the tracking system). The requirements are very realistic and can likely be met with any major ISP/cloud service.
\fi

\iflong
\section{False Positives in the Windows Attack}
\label{appendix:false-positives}
Na\"ively, with an elimination power of $15\cdot G$ bits, $G=2$ should suffice to eliminate almost all false guesses in Phase 2. However, this does not take into account the problem of  false positives  which only differ from the correct key by their few leftmost bits. Below is a detailed analysis of this scenario, specifically for keys differing only in their leftmost bit ($K_{18}$).

\subsection{Correct Counter Value ($\beta$)}
Consider a false key that yields correct data at the end of Phase 1 (i.e. it has the  same $(K_{33},\ldots,K_{62})$ and same $\beta_0 \mod 2^{14}$ as the correct key), but has a different $K_{18}$.

In Section \ref{Section:Phase-2} we saw that for $g$ where $\IP^{g,0}_0 = \IP^0_0$, $K_{18}$ does not affect the expression on the right hand side of eq.~(\ref{eq:phase2-main}). Therefore we can focus on $g$ where $\IP^{g,0}_0 \neq \IP^0_0$ (i.e. $\IP^{g,j}_0 \oplus \IP^0_0=1$). It is still possible in such cases, that the right hand side of eq.~(\ref{eq:phase2-main}) will be identical for the two keys, 
depending on the second from left bit of $T(K,\IP^0 \oplus \IP^{g,j})_{17,\ldots,31} \oplus \Vectorize(\mathit{\IPID}^0-\beta_0 \mod 2^{15})_{17,\ldots,31}$. 

This bit can be written as $(\IP^0_0 \oplus \IP^{g,j}_0)K_{18} \oplus \Psi_1^{g,1}$, where $\Psi_1^{g,j}$ depends on $K_{19},\ldots,K_{62}$, $\mathit{\IPID}^{g,j}$, and $\beta_0 \mod 2^{15}$. Denote the remaining bits of $T(K,\IP^0 \oplus \IP^{g,j})_{17,\ldots,31} \oplus \Vectorize(\mathit{\IPID}^0-\beta_0 \mod 2^{15})_{17,\ldots,31}$ with the corresponding $\Psi^{g,j}_{i-17}$ notation, and note that they do not depend on $K_{18}$ (except for $\Psi^{g,j}_0$, in which $K_{18}$ cancels itself in the subtraction), and hence are identical across the two keys. Therefore the right hand side of eq.~(\ref{eq:phase2-main}) can be written as follows (with $K_{18}$ being in effect only where it is explicitly written):
\begin{eqnarray*}
& 1 & +  \Num(\Psi^{g,1}_0,\Psi^{g,1}_1 \oplus K_{18},\Psi^{g,1}_1,\ldots,\Psi^{g,1}_{14})\\
& - & \Num(\Psi^{g,0}_0,\Psi^{g,0}_1 \oplus K_{18},\Psi^{g,0}_1,\ldots,\Psi^{g,0}_{14}) \mod 2^{15}
\end{eqnarray*}
Now it is easy to see that $K_{18}$ only affects the result if $\Psi^{g,1}_1 \neq \Psi^{g,0}_1$ (i.e. $\Psi^{g,1}_1 \oplus \Psi^{g,0}_1=1$) since only in this case the borrow into the leftmost bit is different.
Assuming $\Psi^{g,1}_1 \oplus \Psi^{g,0}_1$ is random, then a pair of IPs in class $b_g$ with $\IP^{g,0}_0 \neq \IP^0_0$ has a probability $\frac{1}{2}$ to distinguish the two keys. In  other words, there is a lower bound of $$2^{-|\{g|\IP^{g,0}_0 \neq \IP^0_0\}|}=2^{-(G-Q)}$$  on the false positive probability (which due to incorrect keys which differ in $K_{18}$ from the correct key, and whose $\beta_0 \mod 2^{14}$ is correct). 

\subsection{Incorrect Counter Value ($\beta$)}
\label{app:incorrect-beta}
Another source of false positives is the case where Phase 1 emits multiple values for $\beta_0 \mod 2^{14}$ (a correct one, and one or more incorrect ones).
Specifically, consider $\beta_0 \mod 2^{14}$ which differs from the correct value only in its most significant bit. 

If $D^j$ are unchanged, then the correct $K_{33},\ldots,K_{62}$ will solve equations eq.~(\ref{eq:pahse1-main-D}), and thus Phase 1 will emit correct key bits, but incorrect $\beta_0 \mod 2^{14}$ to Phase 2. In order for $D^j$ not to change, the borrow in $\mathit{\IPID}^j-\beta_0\mod 2^{15}$ (ignoring the $j$ addendum since $j$ is small and we are interested in the leftmost bits) must be identical to the borrow in $\mathit{\IPID}^0-\beta_0 \mod 2^{15}$ induced by the change in $\beta_0 \mod 2^{14}$. 

This event only happens when the second-from-left bits in $\IPID^j$ and in $\IPID^0$ are identical. In such a case, Phase 2 will get correct $K_{33},\ldots,K_{62}$, but incorrect $\beta_0 \mod 2^{14}$ (flipped leftmost bit). This happens with probability $2^{-(J-1)}$.

In Phase 2 (eq.~(\ref{eq:phase2-main})), $\Vectorize(\IPID^0-\beta_0 \mod 2^{15})_{17,\ldots,31}$ will have its second-from-left bit flipped. 
Consider now an incorrect key which is identical to the correct key in all bits but $K_{18}$ (which is flipped). The coefficient of $K_{18}$ as linear addendum in the second-from-left bit of $T(K,\IP^0 \oplus \IP^{g,j})_{17,\ldots,31}$ is $(\IP^0 \oplus \IP^{g,j})_0$, so if $(\IP^0 \oplus \IP^{g,j})_0=1$ then the flipping of the second-from-left bit of $\Vectorize(\IPID^0-\beta_0 \mod 2^{15})_{17,\ldots,31}$ is cancelled by the flipping in the second-from-left bit of $T(K,\IP^0 \oplus \IP^{g,j})_{17,\ldots,31}$ induced by the flipped $K_{18}$, and therefore eq.~(\ref{eq:phase2-main}) holds. 

If $(\IP^0 \oplus \IP^{g,j})_0=0$ then the flipping in $\Vectorize(\IPID^0-\beta_0 \mod 2^{15})_{17,\ldots,31}$ may still cancel itself in the subtraction, if $T(K,\IP^0 \oplus \IP^{g,0})_{18}=T(K,\IP^0 \oplus \IP^{g,1})_{18}$, which has $\frac{1}{2}$ probability. 

In conclusion, Phase 2 emits the false key with a probability of approximately (due to neglecting of the $j$ addendum) $$2^{-(J-1)-|\{g|\IP^{g,0}_0 = \IP^0_0\}|}=2^{-(J-1)-Q}.$$

\subsection{Incorrect Data (IP IDs)}
\label{appendix:incorrect-data}
The above discussion assumes that the data fed into Phase 1 (the $\IPID^{g,j}$ values) is correct, i.e. that these are the exact same IP IDs generated by the device.
This assumption may break if one or more of the following conditions hold:
\begin{itemize}
    \itemshort TCP retransmissions combined with lost packets (e.g. a TCP packet is lost, then retransmitted, or a TCP packet is received, but considered lost by the sender, retransmitted, but then the retransmitted packet is lost). In both cases, the receiver has no idea that a packet was lost on the way, so the IP ID counter ``synchornization'' is lost.
    \itemshort The sender (device) sends one or more packets whose IP ID is generated through the same bucket $\beta[i]$, between sending the intended packets to $\IP^0$ and $\IP^{J-1}$ or to $\IP^{g,0}$ and $\IP^{g,1}$. In such case, again there will be a ``synchronization'' loss due to the unaccounted-for increment in $\beta[i]$.
    \itemshort The IP ID is not generated by Windows 8 or later. 
\end{itemize}
We claim that in such cases, with extremely high probability, the algorithm will yield no key at all, which can then be handled by an upper-level logic (e.g. re-test the device to eliminate temporary adverse conditions). Specifically, a case in which the IP ID is not generated by Windows 8 or later is extremely unlikely to yield any key (if the IP ID are randomly or pseudo-randomly generated), so this one is trivially dismissed. One exception is a counter-based IP ID (either global counter, or per-IP counter), which can conform to a key wherein $K_{18}=\cdots=K_{62}=0$ (this key yields $T(K,I)_{17,\cdots,31}=(0,0,\ldots,0)$ for $|I|=32$), but this case can be easily detected since $\IPID^{g,j+1}-\IPID^{g,j} \mod 2^{15}=1$ which is extremely improbable for IP ID's generated by Windows 8 or later.

The remaining cases all involve a situation wherein $\beta[i]$ gets incremented out of sync. Both cases are rare (the first one only happens in adverse network conditions, and in the second one, the probability of a packet destined to a random IP address to hit a specific bucket is $2^{-13}$, and the timing is extremely delicate). Nevertheless, we simulated this condition with 100 randomly chosen sets of 30 IPs ($J=6, G=12, Q=3$), 10,000 tests each, where in each individual test, we chose a random class ($g$) and a random position ($j$) within the class, and incremented $\beta[i]$ once (between calculating $\IPID^{g,j}$ and $\IPID^{g,j+1}$). In each of the 1,000,000 tests, no keys were emitted by the algorithm. 
\fi

\subsection{Optimizing the IP Set for Minimum False Positives}
\label{App:Windows-optimize-FP}
Since (from Table \ref{Table:common-tail}) keys with flipped $K_{18}$ are the source of most false positives, the tracker should choose a set of IP addresses that minimizes (over $Q$) this false positive probability, $2^{-(J-1)-Q}+2^{-(G-Q)}$ 
\iflong
(see Appendix~\ref{appendix:false-positives}.)
\fi
The said minimum is at $Q=\nicefrac{G-(J-1)}{2}$, and yields false positive leading term of $2 \cdot 2^{-\frac{G+J-1}{2}}$.
For $G=12$, $J=6$, the optimum is at 
\iflong
$Q=3$ or $Q=4$, i.e. there are 3 (or 4) $b_g$ values whose leftmost bit is identical to the one of $b_0$. 
In such case, the combined probability of a false key which differs in $K_{18}$ is $2^{-8}+2^{-9}=0.005859375$. 
Note that in Table \ref{Table:common-tail}, the probability of a common tail length of 44 bits is exactly the probability of a false key differing in $K_{18}$ only, and the value measured, 0.0058328, is in line with the theory ($E=0.005859375$, for 10,000,000 tests, $\sigma= 2.41\times 10^{-5}$, so $0.0058328=E-1.1\sigma$).
\fi

\iflong
\section{Windows: Exposing Kernel Data}
\label{appendix:exposing-kernel-data}
Prior to Microsoft's October 2018 Security Update, The $\beta$ array ({\tt IpFragmentIdIncrementTable} in the Windows {\tt tcpip.sys} module) was not initialized (except for the first two cells), so it contained data left over in memory from the previous kernel modules. This data happens to be an {\em extension} of the data structure 
{\tt RTL\_PROCESS\_MODULES} \cite{RTL-PROCESS-MODULES} which is essentially a list of {\tt RTL\_PROCESS\_MODULE\_INFORMATION} structures \cite{RTL-PROCESS-MODULES-INFORMATION}, each describing a loaded kernel module.
Knowing $K$, the attacker forces the device to send a packet to 8192 IP addresses, chosen so that $T(K,\IP^{g}_{0,\ldots,15})_{19,\ldots,31}$ takes all possible $M=8192$ values (note this requires a stronger adversary model a-la MitM, and that the attacker can only choose the IP set after $K$ is reconstructed). This in turn ensures that $i_g$ goes through all possible $M=8192$ values. Per Section \ref{Section:Phase-2}, the attacker extracts $\beta[i_g] \mod 2^{14}$ for each packet. However, the attacker still does not know the order, i.e. which index $i$ belongs to which value extracted. The attacker now enumerates over $(K2 \oplus T(K,\IP_{SRC}))_{19,\ldots,31}$ (13 bits), and together with the (known) $T(K,\IP^{g}_{0,\ldots,15})_{19,\ldots,31}$, calculates $i$ for each $\IP^{g}$ and $\IPID^{g}$. Given the data structure of {\tt RTL\_PROCESS\_MODULES}, and that the first module listed is always {\tt \textbackslash SystemRoot\textbackslash system32\textbackslash ntoskrnl.exe}, the attacker can eliminate false guesses and obtain the kernel data in the correct order (15 least significant bits from each 32-bit {\tt DWORD}). Note that {\tt RTL\_PROCESS\_MODULE\_INFORMATION} contains kernel address data, so this method partially bypasses KASLR.
\fi

\iflong
\section{Attacking RFC 7739's Hash-Based Fragment Identification Selection Algorithm}
\label{appendix:attacking-crypto-T}
The Windows algorithm is an implementation of the following abstract scheme (as proposed in RFC 7739 \cite[Section 5.3]{rfc7739}):
\begin{multline*}
IPID = F(IP_{SRC}, IP_{DST}, Key_1)  + \\
counter[G(IP_{SRC}, ClassB(IP_{DST}), Key_2)] \mod 2^{15}
\end{multline*}
This scheme lends itself easily to tracking 
as long as $\IP_{SRC}$ does not change (i.e. the device remains in the same network). Observe that packets sent to IPs in the same class B network have their IP IDs generated using the same counter. The tracker needs to control several ($J$) IPs {\em in the same class B network} - denote these by $IP^j$ (where $0 \leq j < J$), and force the device to send packets to these addresses in rapid succession, observing their corresponding IP ID values $\IPID^j$. Observe that 
\begin{multline*}
\IPID^j-\IPID^0 \mod 2^{15}=\\
j+F(IP_{SRC}, IP^j, key_1)-F(IP_{SRC}, IP^0, key_1) \mod 2^{15}
\end{multline*}
And the right hand side is fixed (for fixed $\IP_{SRC}$).
The device ID then is simply a vector of $\IPID^j-\IPID^0 \mod 2^{15}$ ($J-1$ values). 
Since each element in the vector is 15 bit, 3-4 elements suffice to fingerprint devices in a large scale deployment, hence $J=4$ or $J=5$ IP addresses (in the same class B network) are needed.
\fi

\iflong
\section{Analysis of the Effective Key Space in Attacking Algorithm \it A\textsubscript{3}}
\label{app:netspace}
The function $g(net)$ is a shift right of $net$ by $\rho$ (architecture dependent) bits.\footnote{$\rho={\tt L1\_CACHE\_SHIFT}$ (\url{https://git.kernel.org/pub/scm/linux/kernel/git/stable/linux-stable.git/plain/include/net/netns/hash.h}) in kernel versions up to 4.18, $\rho=\lfloor\log_2 {\tt sizeof(struct \; net)}\rfloor$ in versions 4.19 and later.

{\tt L1\_CACHE\_SHIFT} is $\log_2$ of the L1 cache line size, and is architecture/chipset dependent. In x64 architecture (modern Linux) and in older ARM64 chipsets (e.g. Snapdragon 835) it is 6, and in newer ARM64 chipsets (e.g. Snapdragon 845) it is 7. $\rho$ can be extracted statically from the kernel image file.

{\tt sizeof(struct net)} is determined by many configuration options, and therefore can considerably vary among builds and architectures. In one Azure Ubuntu 16.04.1 (Linux kernel 4.15.0) machine with x64 architecture, we observed a value of 6080 bytes, so we assume most values will be in the range 4096-8191, i.e.  $\lfloor\log_2 {\tt sizeof(struct \; net)}\rfloor=12$.}
{\tt net} is the address of the network namespace object associated with the current container (OS isolation compartment). All such objects are maintained by Linux in a linked list. The head of the list is a global variable ({\tt init\_net}),\footnote{\url{https://elixir.bootlin.com/linux/v4.0/source/net/core/net_namespace.c}} which means that the {\em first} container will have its {\tt net} address in the {\tt .data} segment of the kernel image, i.e. will have a fixed offset from the start of the kernel image in memory. This is relevant for all single-container Linux installations: desktops/laptops, most servers, and Android devices.
This offset from the start of the kernel image is determined in compile time, and is build-specific. Given the build, it is possible to determine the offset  offline, since it depends only on the Linux build, rather than on the hardware or any temporal conditions, and can be extracted from the kernel image file (indeed, we wrote a Perl script that extracts it from some Linux and Android kernel image files). Also, from the kernel image file it is also possible to extract the value of  $\lfloor\log_2 {\tt sizeof(struct \; net)}\rfloor$ if needed (e.g. from the implementation of {\tt net\_hash\_mix}.)\footnote{\url{https://git.kernel.org/pub/scm/linux/kernel/git/stable/linux-stable.git/plain/include/net/netns/hash.h}}

The build identity can be either inferred or obtained in many ways. In many Android devices for example, the Chrome browser reports the exact build identifier as part of the {\tt User-Agent} HTTP request header. And in Linux, when the distribution version is known, there are only a few different builds that are associated with it. Once the build identity is known, it is possible (in Linux) to obtain the kernel image file directly via a download site, or (in Android) to extract the kernel image from a stock ROM file from a download site.
Henceforth we assume that the offset of {\tt init\_net} (from the kernel image base address) is known.

As for the Android/ARM64: the ``official'' KASLR support for ARM64 architecture was introduced in Linux kernel 4.6. Google back-ported it to kernel 4.4.56 used in Android 8.x \cite{android-kaslr}, Android builds based on kernel 4.4.56+ include KASLR, and Androids based on earlier builds do not. 

\subsection{Linux (x86\_64 Architecture), KASLR Disabled}
In some Linux distributions, KASLR is disabled by default (e.g.,  in older Ubuntu distributions.)\footnote{\url{https://wiki.ubuntu.com/Security/Features}}
Without KASLR, the kernel image is always loaded at virtual address\\ {\tt \_\_START\_KERNEL=\_\_START\_KERNEL\_map + \_\_PHYSICAL\_START=0xffffffff80000000+\\
CONFIG\_PHYSICAL\_START},\footnote{ \url{https://elixir.bootlin.com/linux/v4.0/source/arch/x86/include/asm/page\_types.h}, \url{https://elixir.bootlin.com/linux/v4.8/source/arch/x86/include/asm/page\_64\_types.h}} and by default, {\tt CONFIG\_PHYSICAL\_START=0x1000000},\footnote{\url{https://elixir.bootlin.com/linux/v4.0/source/arch/x86/Kconfig}} thus {\tt net} will be {\tt 0xffffffff81000000} plus the {\tt init\_net} offset.
In this case, $|\{g(net)\}|=1$, thus $|W|=2^{32}$. 
In our measurements, given an average (based on simulations) of 70.45 pairs (of which 37.48 pairs on average are correct), Azure's second weakest and cheapest VM (B1ms class, 1 vCPU \cite{Azure-VMs}) can go over all $2^{32}$ keys in 2077 seconds (less than 35 minutes). A machine/cluster of 128 CPUs can provide real-time performance in this case. 

\subsection{Linux (x86\_64 Architecture), KASLR Enabled}
When KASLR is enabled, a random quantity is added to the original kernel image base address ({\tt 0xffffffff81000000}), and the new value is used as the effective image base address. 
This quantity is computed in {\tt find\_random\_virt\_addr}\footnote{\url{https://elixir.bootlin.com/linux/v4.8/source/arch/x86/boot/compressed/kaslr.c}} as a subset of all possible increments of {\tt CONFIG\_PHYSICAL\_ALIGN}=2MB\footnote{\url{https://elixir.bootlin.com/linux/v4.0/source/arch/x86/Kconfig}} from 0 to {\tt KERNEL\_IMAGE\_SIZE}=1GB,\footnote{\url{https://elixir.bootlin.com/linux/v4.8/source/arch/x86/include/asm/page\_64\_types.h}} thus there are up to 512 possible values. Note that all the random bits are part of the least significant 30 bits of the address, and therefore all of them are part of $g(net)$, regardless of the value of $\rho$. Hence $|\{g(net)\}|=2^9$ and $|W|=2^{41}$. 

This is still a realistic scenario for an attack, possibly even in real time given a powerful enough CPU (or CPUs - the exhaustive search can be extremely parallelized). For example, we estimate that 10,000 B1s virtual machines in Microsoft Azure (each \$3.20/month) would be able to enumerate this key space (for an average 70.45 pairs of IP addresses) in around 99 seconds. Using stronger machines can bring this down to real-time performance. 
Given $g(net)$ it is possible to reconstruct {\tt init\_net}, and subtracting its offset, to get to {\tt \_\_START\_KERNEL}, from which it is easy to find all addresses of code and global/static variables in the kernel. Therefore the technique provides a full kernel KASLR bypass.

\subsection{Android (ARM 64-Bit Architecture), KASLR Disabled}
The original (non-randomized) kernel image base address is ${\tt KIMAGE\_VADDR}=2^{64}-2^{\tt CONFIG\_ARM64\_VA\_BITS}+2^{27}$, with the {\tt .text} segment {\tt TEXT\_OFFSET=0x80000} above that.\footnote{\url{https://elixir.bootlin.com/linux/v4.6/source/arch/arm64/include/asm/memory.h}, \url{https://elixir.bootlin.com/linux/v4.6/source/arch/arm64/Makefile}, \url{https://elixir.bootlin.com/linux/v4.6/source/drivers/firmware/efi/libstub/arm64-stub.c}}
and {\tt CONFIG\_ARM64\_VA\_BITS=39} by default.\footnote{\url{https://elixir.bootlin.com/linux/v4.6/source/arch/arm64/Kconfig}} All the Android kernel images we reviewed use these default values.
The kernel image {\tt .text} segment start address is therefore always {\tt 0xffffff8008080000}, and {\tt init\_net} takes its offset from this address. Since there is no KASLR, the offset of {\tt init\_net} is constant, hence $|\{g(net)\}|=1$ and $|W|=2^{32}$. Practically, we are not aware of Android kernels wherein {\tt CONFIG\_NET\_NS} is enabled, and KASLR is not enabled.
\subsection{Android (ARM 64-Bit Architecture), KASLR Enabled}

Just like in the x86\_64 architecture, a random quantity is added to the original kernel image base address. The quantity is computed in {\tt kaslr\_early\_init} as {\tt ARM64\_VA\_BITS}-23 random bits followed by 21 zero bits.\footnote{\url{https://elixir.bootlin.com/linux/v4.6/source/arch/arm64/kernel/kaslr.c}} Thus the number of random bits is 16. Since $g(net)$ is truncated to 32 bits (after a right shift of {\tt init\_net} by $\rho$, where $\rho={\tt L1\_CACHE\_SHIFT}=6$ or $\rho=7$ for kernel version 4.18.x and earlier, and $\rho=\lfloor\log_2 {\tt sizeof(net)}\rfloor \approx 12$ for kernel version 4.19 and later), all 16 random bits will be part of $g(net)$.Therefore $|\{g(net)\}|=2^{16}$ and $|W|=2^{48}$. While not economically efficient, it is still possible to amass enough computing power to extract this data in specific cases. 
Given $g(net)$ it is possible to reconstruct {\tt init\_net}, and subtracting its offset, to get to {\tt KIMAGE\_VADDR}, from which it is easy to find all addresses of code and global/static variables in the kernel. Therefore the technique provides a full KASLR bypass.\\
NOTE: Samsung devices implement a different KASLR mechanism (part of Samsung Knox). 
The implementation details are not public, but assuming KASLR is implemented as an offset of up to 16 entropy bits, our analysis of Android applies for Samsung devices as well.

\fi

\iflong
\section{Other Linux/Android Attack Scenarios}
\label{Appendix:other-attacks}
\subsection{KASLR Attacks}
\label{Appendix:non-browser-kaslr-attacks}
In previous sections, we focused on an attack scenario wherein a client device accesses a web page via a browser. This scenario is relevant for both tracking and KASLR attacks. There are, however, additional attack scenarios which are relevant to (only) the KASLR attack.

It should be noted that the attack technique can be used with various protocols over UDP (not just WebRTC/STUN) - particularly with DNS. Also, non-UDP stateless protocols can be used, e.g. ICMP. In all such cases, the attacker needs to know the internal IP address of the target device (in WebRTC this can be obtained as part of the protocol) -- either via prior knowledge or via additional enumeration (Appendix~\ref{Appendix:internal-ip-disclosure}). 

Based on this, the KASLR attack can be applied also to servers, and to client devices in additional scenarios:
\begin{itemize}
    \item Attacking UDP-based servers: the attacker can send queries from multiple IP addresses to the target server, and collect answers (which are sent over UDP) and analyze their IP IDs. This applies to any UDP-based protocol, e.g. DNS. The attacker can also force a UDP-based server to send outgoing packets to attacker IPs (without sending packets to the server from these IPs), e.g. in a DNS resolver scenario, the attacker can provide the resolver with an authoritative answer containing multiple NS records pointing at the attacker's hosts.
    \item ICMP Echo Request+Reply packets (ping): the attacker can send ICMP Echo Request packets from multiple IP addresses to the target device (client or server), and assuming the ICMP packets are not dropped/rejected (as often happens), the device will return ICMP Echo Reply packets, which (ICMP being a stateless protocol over IP, hence the analysis applies to its IP ID) can be used to attack KASLR.
\end{itemize}

\subsection{Internal IP Disclosure}
\label{Appendix:internal-ip-disclosure}
In order to extract $key$, the attacker needs to know the (internal) IP address used as a source address for hashing. When the attack is browser-based, WebRTC can be used to extract the internal IP address \cite{webrtc-attack}. This issue is well known, and a standartization effort around remediation is in the works (currently an RFC draft \cite{I-D.ietf-rtcweb-mdns-ice-candidates}). In accordance with this draft RFC, Google introduced a flag {\tt enable-webrtc-hide-local-ips-with-mdns} to Google Chrome v72 to prevent this leakage \cite{chrome-v72-webrtc-release-notes}. This flag is turned off by default, and is not available for Chrome in Android (it is available for Chrome in Linux). As such, at present, it does not have any significant impact on our attack. 

When WebRTC is not available (e.g. the attack is not browser-based, as in Appendix~\ref{Appendix:non-browser-kaslr-attacks}), or when WebRTC cannot be used to disclose the internal IP address (e.g. Linux Chrome users who explicitly turn on {\tt enable-webrtc-hide-local-ips-with-mdns}), the attacker can combine the enumeration over $key$ with enumerartion over the internal IP address space, assuming prior knowledge of that space. For example, if it is known that the internal IP is in the 192.168.0.0/16 address space, then this adds 16 bits to the enumeration of $key$, which is still feasible for a 32-bit $key$. This yields both $key$ and the internal IP address. Another example is Microsoft Azure, which assigns internal IP addresses to servers in a customer farm sequentially, starting from 10.0.1.1. Therefore, enumerating according to this order will require $\log_2{F}$ additional bits, where $F$ is the farm size.

Disclosing the internal IP address of a server is considered a vulnerability in itself. Specifically, it is a violation of the PCI DSS version 3.2.1 article 1.3.7 (``Do not disclose private IP addresses and routing information to unauthorized parties'') \cite{pci-dss}.
\fi

\end{document}